
%

\def\answ{b }

%
%
%
%
%
\def\unredoffs{\hoffset-.14truein\voffset-.2truein} 
\def\redoffs{\voffset=-.45truein\hoffset=-.21truein} 
\def\speclscape{}
%
%
\newbox\leftpage \newdimen\fullhsize \newdimen\hstitle \newdimen\hsbody
\tolerance=1000\hfuzz=2pt
\catcode`\@=11 
\def\bigans{b }
%
\ifx\answ\bigans\message{(This will come out unreduced.}
\magnification=1000\unredoffs\baselineskip=16pt plus 2pt minus 1pt
\hsbody=\hsize \hstitle=\hsize 
\else\message{(This will be reduced.} \let\l@r=L
\magnification=1000\baselineskip=16pt plus 2pt minus 1pt \vsize=7truein
\redoffs \hstitle=8truein\hsbody=4.75truein\fullhsize=10truein\hsize=\hsbody
\output={\ifnum\pageno=0 
  \shipout\vbox{\speclscape{\hsize\fullhsize\makeheadline}
    \hbox to \fullhsize{\hfill\pagebody\hfill}}\advancepageno
  \else
  \almostshipout{\leftline{\vbox{\pagebody\makefootline}}}\advancepageno 
  \fi}
\def\almostshipout#1{\if L\l@r \count1=1 \message{[\the\count0.\the\count1]}
      \global\setbox\leftpage=#1 \global\let\l@r=R
 \else \count1=2
  \shipout\vbox{\speclscape{\hsize\fullhsize\makeheadline}
      \hbox to\fullhsize{\box\leftpage\hfil#1}}  \global\let\l@r=L\fi}
\fi
%
\newcount\yearltd\yearltd=\year\advance\yearltd by -1900

%
%

\def\draftmode{\message{ DRAFTMODE }\def\draftdate{{\rm preliminary draft:
\number\month/\number\day/\number\yearltd\ \ \hourmin}}%
\headline={\hfil\draftdate}\writelabels\baselineskip=20pt plus 2pt minus 2pt
 {\count255=\time\divide\count255 by 60 \xdef\hourmin{\number\count255}
  \multiply\count255 by-60\advance\count255 by\time
  \xdef\hourmin{\hourmin:\ifnum\count255<10 0\fi\the\count255}}}
\def\nolabels{\def\wrlabeL##1{}\def\eqlabeL##1{}\def\reflabeL##1{}}
\def\writelabels{\def\wrlabeL##1{\leavevmode\vadjust{\rlap{\smash%
{\line{{\escapechar=` \hfill\rlap{\sevenrm\hskip.03in\string##1}}}}}}}%
\def\eqlabeL##1{{\escapechar-1\rlap{\sevenrm\hskip.05in\string##1}}}%
\def\reflabeL##1{\noexpand\llap{\noexpand\sevenrm\string\string\string##1}}}
\nolabels
%
\global\newcount\secno \global\secno=0
\global\newcount\meqno \global\meqno=1
\def\newsec#1{\global\advance\secno by1\message{(\the\secno. #1)}
\global\subsecno=0\eqnres@t\noindent{\bf\the\secno. #1}
\writetoca{{\secsym} {#1}}\par\nobreak\medskip\nobreak}
\def\eqnres@t{\xdef\secsym{\the\secno.}\global\meqno=1\bigbreak\bigskip}
\def\sequentialequations{\def\eqnres@t{\bigbreak}}\xdef\secsym{}
\global\newcount\subsecno \global\subsecno=0
\def\subsec#1{\global\advance\subsecno by1\message{(\secsym\the\subsecno. #1)}
\ifnum\lastpenalty>9000\else\bigbreak\fi
\noindent{\it\secsym\the\subsecno. #1}\writetoca{\string\quad 
{\secsym\the\subsecno.} {#1}}\par\nobreak\medskip\nobreak}
\def\appendix#1#2{\global\meqno=1\global\subsecno=0\xdef\secsym{\hbox{#1.}}
\bigbreak\bigskip\noindent{\bf Appendix #1. #2}\message{(#1. #2)}
\writetoca{Appendix {#1.} {#2}}\par\nobreak\medskip\nobreak}
%
%
\def\eqnn#1{\xdef #1{(\secsym\the\meqno)}\writedef{#1\leftbracket#1}%
\global\advance\meqno by1\wrlabeL#1}
\def\eqna#1{\xdef #1##1{\hbox{$(\secsym\the\meqno##1)$}}
\writedef{#1\numbersign1\leftbracket#1{\numbersign1}}%
\global\advance\meqno by1\wrlabeL{#1$\{\}$}}
\def\eqn#1#2{\xdef #1{(\secsym\the\meqno)}\writedef{#1\leftbracket#1}%
\global\advance\meqno by1$$#2\eqno#1\eqlabeL#1$$}
%
\newskip\footskip\footskip14pt plus 1pt minus 1pt 
\def\footnotefont{\ninepoint}\def\f@t#1{\footnotefont #1\@foot}
\def\f@@t{\baselineskip\footskip\bgroup\footnotefont\aftergroup\@foot\let\next}
\setbox\strutbox=\hbox{\vrule height9.5pt depth4.5pt width0pt}
\global\newcount\ftno \global\ftno=0
\def\foot{\global\advance\ftno by1\footnote{$^{\the\ftno}$}}
%
\newwrite\ftfile   
\def\footend{\def\foot{\global\advance\ftno by1\chardef\wfile=\ftfile
$^{\the\ftno}$\ifnum\ftno=1\immediate\openout\ftfile=foots.tmp\fi%
\immediate\write\ftfile{\noexpand\smallskip%
\noexpand\item{f\the\ftno:\ }\pctsign}\findarg}%
\def\footatend{\vfill\eject\immediate\closeout\ftfile{\parindent=20pt
\centerline{\bf Footnotes}\nobreak\bigskip\input foots.tmp }}}
\def\footatend{}
%
%
\global\newcount\refno \global\refno=1
\newwrite\rfile
\def\ref{[\the\refno]\nref}
\def\nref#1{\xdef#1{[\the\refno]}\writedef{#1\leftbracket#1}%
\ifnum\refno=1\immediate\openout\rfile=refs.tmp\fi
\global\advance\refno by1\chardef\wfile=\rfile\immediate
\write\rfile{\noexpand\item{#1\ }\reflabeL{#1\hskip.31in}\pctsign}\findarg}
\def\findarg#1#{\begingroup\obeylines\newlinechar=`\^^M\pass@rg}
{\obeylines\gdef\pass@rg#1{\writ@line\relax #1^^M\hbox{}^^M}%
\gdef\writ@line#1^^M{\expandafter\toks0\expandafter{\striprel@x #1}%
\edef\next{\the\toks0}\ifx\next\em@rk\let\next=\endgroup\else\ifx\next\empty%
\else\immediate\write\wfile{\the\toks0}\fi\let\next=\writ@line\fi\next\relax}}
\def\striprel@x#1{} \def\em@rk{\hbox{}} 
\def\lref{\begingroup\obeylines\lr@f}
\def\lr@f#1#2{\gdef#1{\ref#1{#2}}\endgroup\unskip}
\def\semi{;\hfil\break}
\def\addref#1{\immediate\write\rfile{\noexpand\item{}#1}} 
\def\startrefs#1{\immediate\openout\rfile=refs.tmp\refno=#1}
\def\xref{\expandafter\xr@f}\def\xr@f[#1]{#1}
\def\refs#1{\count255=1[\r@fs #1{\hbox{}}]}
\def\r@fs#1{\ifx\und@fined#1\message{reflabel \string#1 is undefined.}%
\nref#1{need to supply reference \string#1.}\fi%
\vphantom{\hphantom{#1}}\edef\next{#1}\ifx\next\em@rk\def\next{}%
\else\ifx\next#1\ifodd\count255\relax\xref#1\count255=0\fi%
\else#1\count255=1\fi\let\next=\r@fs\fi\next}
%

%
\newwrite\ffile\global\newcount\figno \global\figno=1
\def\fig{fig.~\the\figno\nfig}
\def\nfig#1{\xdef#1{fig.~\the\figno}%
\writedef{#1\leftbracket fig.\noexpand~\the\figno}%
\ifnum\figno=1\immediate\openout\ffile=figs.tmp\fi\chardef\wfile=\ffile%
\immediate\write\ffile{\noexpand\medskip\noexpand\item{Fig.\ \the\figno. }
\reflabeL{#1\hskip.55in}\pctsign}\global\advance\figno by1\findarg}
\def\xfig{\expandafter\xf@g}\def\xf@g fig.\penalty\@M\ {}
\def\figs#1{figs.~\f@gs #1{\hbox{}}}
\def\f@gs#1{\edef\next{#1}\ifx\next\em@rk\def\next{}\else
\ifx\next#1\xfig #1\else#1\fi\let\next=\f@gs\fi\next}
\newwrite\lfile
{\escapechar-1\xdef\pctsign{\string\%}\xdef\leftbracket{\string\{}
\xdef\rightbracket{\string\}}\xdef\numbersign{\string\#}}
\def\writedefs{\immediate\openout\lfile=labeldefs.tmp \def\writedef##1{%
\immediate\write\lfile{\string\def\string##1\rightbracket}}}
\def\writestop{\def\writestoppt{\immediate\write\lfile{\string\pageno%
\the\pageno\string\startrefs\leftbracket\the\refno\rightbracket%
\string\def\string\secsym\leftbracket\secsym\rightbracket%
\string\secno\the\secno\string\meqno\the\meqno}\immediate\closeout\lfile}}
\def\writestoppt{}\def\writedef#1{}
\def\seclab#1{\xdef #1{\the\secno}\writedef{#1\leftbracket#1}\wrlabeL{#1=#1}}
\def\subseclab#1{\xdef #1{\secsym\the\subsecno}%
\writedef{#1\leftbracket#1}\wrlabeL{#1=#1}}
\newwrite\tfile \def\writetoca#1{}
\def\leaderfill{\leaders\hbox to 1em{\hss.\hss}\hfill}
\def\writetoc{\immediate\openout\tfile=toc.tmp 
   \def\writetoca##1{{\edef\next{\write\tfile{\noindent ##1 
   \string\leaderfill {\noexpand\number\pageno} \par}}\next}}}

%
\catcode`\@=12 
%
\edef\tfontsize{\ifx\answ\bigans scaled\magstep3\else scaled\magstep4\fi}
 \tfontsize  \tfontsize
 \tfontsize \font\titlei=cmmi10 \tfontsize
\font\titleis=cmmi7 \tfontsize \font\titleiss=cmmi5 \tfontsize
\font\titlesy=cmsy10 \tfontsize \font\titlesys=cmsy7 \tfontsize
\font\titlesyss=cmsy5 \tfontsize  \tfontsize
\skewchar\titlei='177 \skewchar\titleis='177 \skewchar\titleiss='177
\skewchar\titlesy='60 \skewchar\titlesys='60 \skewchar\titlesyss='60
 \ifx\answ\bigans\else scaled\magstep1\fi
\ifx\answ\bigans\else

 \font\absi=cmmi10 scaled\magstep1
\font\absis=cmmi7 scaled\magstep1 \font\absiss=cmmi5 scaled\magstep1
\font\abssy=cmsy10 scaled\magstep1 \font\abssys=cmsy7 scaled\magstep1
\font\abssyss=cmsy5 scaled\magstep1 
\skewchar\absi='177 \skewchar\absis='177 \skewchar\absiss='177
\skewchar\abssy='60 \skewchar\abssys='60 \skewchar\abssyss='60
\fi
\def\tenpoint{\def\rm{\fam0\tenrm}
\textfont0=\tenrm \scriptfont0=\sevenrm \scriptscriptfont0=\fiverm
\textfont1=\teni  \scriptfont1=\seveni  \scriptscriptfont1=\fivei
\textfont2=\tensy \scriptfont2=\sevensy \scriptscriptfont2=\fivesy
\textfont\itfam=\tenit \def\it{\fam\itfam\tenit}\def\footnotefont{\ninepoint}%
\textfont\bffam=\tenbf \def\bf{\fam\bffam\tenbf}\def\sl{\fam\slfam\tensl}\rm}
\font\ninerm=cmr9 \font\sixrm=cmr6 \font\ninei=cmmi9 \font\sixi=cmmi6 
\font\ninesy=cmsy9 \font\sixsy=cmsy6 \font\ninebf=cmbx9 
\font\nineit=cmti9 \font\ninesl=cmsl9 \skewchar\ninei='177
\skewchar\sixi='177 \skewchar\ninesy='60 \skewchar\sixsy='60 
\def\ninepoint{\def\rm{\fam0\ninerm}
\textfont0=\ninerm \scriptfont0=\sixrm \scriptscriptfont0=\fiverm
\textfont1=\ninei \scriptfont1=\sixi \scriptscriptfont1=\fivei
\textfont2=\ninesy \scriptfont2=\sixsy \scriptscriptfont2=\fivesy
\textfont\itfam=\ninei \def\it{\fam\itfam\nineit}\def\sl{\fam\slfam\ninesl}%
\textfont\bffam=\ninebf \def\bf{\fam\bffam\ninebf}\rm} 
%
%

\hyphenation{anom-aly anom-alies coun-ter-term coun-ter-terms}
\def\inv{^{\raise.15ex\hbox{${\scriptscriptstyle -}$}\kern-.05em 1}}

\def\Dsl{\,\raise.15ex\hbox{/}\mkern-13.5mu D} 
\def\dsl{\raise.15ex\hbox{/}\kern-.57em\partial}

\def\tr{{\rm tr}} 

\def\lspace{\ifx\answ\bigans{}\else\qquad\fi}
\def\lbspace{\ifx\answ\bigans{}\else\hskip-.2in\fi} 
\def\boxeqn#1{\vcenter{\vbox{\hrule\hbox{\vrule\kern3pt\vbox{\kern3pt
	\hbox{${\displaystyle #1}$}\kern3pt}\kern3pt\vrule}\hrule}}}
\def\mbox#1#2{\vcenter{\hrule \hbox{\vrule height#2in
		\kern#1in \vrule} \hrule}}  
%
\def\CA{{\cal A}}    
 \def\CH{{\cal H}} \def\CI{{\cal I}}

\def\darr#1{\raise1.5ex\hbox{$\leftrightarrow$}\mkern-16.5mu #1}

\def\roughly#1{\raise.3ex\hbox{$#1$\kern-.75em\lower1ex\hbox{$\sim$}}}

\input labeldefs.tmp
\writedefs
\overfullrule=0pt

\def\omit#1{{}}
\def\omitt#1{{}}
\def\omitt#1{{#1}}

\input epsf

\def\fig#1#2#3{
\xdef#1{\the\figno}
\writedef{#1\leftbracket \the\figno}
\nobreak
\par\begingroup\parindent=0pt\leftskip=1cm\rightskip=1cm\parindent=0pt
\baselineskip=11pt
\midinsert
\centerline{#3}
\vskip 12pt
{\bf Fig. \the\figno:} #2\par
\endinsert\endgroup\par
\goodbreak
\global\advance\figno by1
}
\newwrite\tfile\global\newcount\tabno \global\tabno=1
\def\tab#1#2#3{
\xdef#1{\the\tabno}
\writedef{#1\leftbracket \the\tabno}
\nobreak
\par\begingroup\parindent=0pt\leftskip=1cm\rightskip=1cm\parindent=0pt
\baselineskip=11pt
\midinsert
\centerline{#3}
\vskip 12pt
{\bf Tab. \the\tabno:} #2\par
\endinsert\endgroup\par
\goodbreak
\global\advance\tabno by1
}
%

%

%
%
%
%



\def\oh{{1\over 2}}
\def\bz{\bar z}
\def\bra{\langle}\def\ket{\rangle}

\def\nind{\par\noindent}

%


\input epsf.tex

\font\eightrm=cmr8\font\eighti=cmmi8
\font\eightsy=cmsy8\font\eightit=cmti8
\font\eightsl=cmsl8\font\eighttt=cmtt8\font\eightbf=cmbx8
\font\sixrm=cmr6\font\sixi=cmmi6
\font\sixsy=cmsy6

\font\sixbf=cmbx6

\def\tenpoint{%
\textfont0=\tenrm \scriptfont0=\sevenrm
\scriptscriptfont0=\fiverm \def\rm{\fam0\tenrm}%
\textfont1=\teni \scriptfont1=\seveni
\scriptscriptfont1=\fivei \def\oldstyle{\fam1\teni}%
\textfont2=\tensy \scriptfont2=\sevensy
\scriptscriptfont2=\fivesy
\textfont\itfam=\tenit \def\it{\fam\itfam\tenit}%
\textfont\slfam=\tensl \def\sl{\fam\slfam\tensl}%
\textfont\ttfam=\tentt \def\tt{\fam\ttfam\tentt}%
\textfont\bffam=\tenbf \scriptfont\bffam=\sevenbf
\scriptscriptfont\bffam=\fivebf \def\bf{\fam\bffam\tenbf}%
\abovedisplayskip=12pt plus 3pt minus 9pt
\belowdisplayskip=\abovedisplayskip
\abovedisplayshortskip=0pt plus 3pt 
\belowdisplayshortskip=7pt plus 3pt minus 4pt
\smallskipamount=3pt plus 1pt minus 1pt
\medskipamount=6pt plus 2pt minus 2pt
\bigskipamount=12pt plus 4pt minus 4pt
\setbox\strutbox=\hbox{\vrule height8.5pt depth3.5pt width 0pt} %
\normalbaselineskip=12pt \normalbaselines
\rm}

\def\eightpoint{%
\textfont0=\eightrm \scriptfont0=\sixrm
\scriptscriptfont0=\fiverm \def\rm{\fam0\eightrm}%
\textfont1=\eighti \scriptfont1=\sixi
\scriptscriptfont1=\fivei \def\oldstyle{\fam1\eighti}%
\textfont2=\eightsy 
\scriptscriptfont2=\fivesy
\textfont\itfam=\eightit \def\it{\fam\itfam\eightit}%
\textfont\slfam=\eightsl \def\sl{\fam\slfam\eightsl}%
\textfont\ttfam=\eighttt \def\tt{\fam\ttfam\eighttt}%
\textfont\bffam=\eightbf \scriptfont\bffam=\sixbf
\scriptscriptfont\bffam=\fivebf \def\bf{\fam\bffam\eightbf}%
\abovedisplayskip=9pt plus 2pt minus 6pt
\belowdisplayskip=\abovedisplayskip
\abovedisplayshortskip=0pt plus 2pt 
\belowdisplayshortskip=5pt plus 2pt minus 3pt
\smallskipamount=2pt plus 1pt minus 1pt
\medskipamount=4pt plus 2pt minus 2pt
\bigskipamount=9pt plus 4pt minus 4pt
\setbox\strutbox=\hbox{\vrule height7pt depth2pt width 0pt} %
\normalbaselineskip=9pt \normalbaselines
\rm}

\def\petit{\vskip3mm\eightpoint \skewchar\eighti='177 \skewchar\sixi='177 
\skewchar\eightsy='60 \skewchar\sixsy='60}


\newcount\figno
\figno=0
\def\fig#1#2#3{
\par\begingroup\parindent=0pt\leftskip=1cm\rightskip=1cm\parindent=0pt
\baselineskip=11pt
\global\advance\figno by 1
\midinsert
\epsfxsize=#3
\centerline{\epsfbox{#2}}
\vskip 12pt
{\eightbf Fig. \the\figno:} {\eightrm #1}\par
\endinsert\endgroup\par
}
\def\figlabel#1{\xdef#1{\the\figno}}

\def\encadremath#1{\vbox{\hrule\hbox{\vrule\kern4pt\vbox{\kern4pt
\hbox{$\displaystyle #1$}\kern4pt}
\kern4pt\vrule}\hrule}}

\newwrite\tfile\global\newcount\tabno \global\tabno=1
\def\tab#1#2#3{
\xdef#1{\the\tabno}
\writedef{#1\leftbracket \the\tabno}
\nobreak
\par\begingroup\parindent=0pt\leftskip=1cm\rightskip=1cm\parindent=0pt
\baselineskip=11pt
\midinsert
\centerline{#3}
\vskip 12pt
{\bf Tab. \the\tabno:} #2\par
\endinsert\endgroup\par
\goodbreak
\global\advance\tabno by1
}

%
%
\def\za{\alpha} \def\zb{\beta} \def\zg{\gamma} \def\zd{\delta}

\def\IZ{Z\!\!\!Z}
\def\dC{C\kern-6.5pt I}

\def\Exp{{\rm Exp}}
\def\Fo{{{}^{(1)}\!F}}
\def\N{Z}

\def\Ex{{\bf Exercise~}}

\def\bi{\bar i}\def\bj{\bar j}\def\jb{\bar j}
 \def\bz{\bar z}
\def\bn{{\bf n}}
\def\bbn{\bar{\bf n}}

\def\bL{\bar L}

\def\tn{\tilde n}
\def\tq{\tilde q}
\def\tV{{\tilde V}}
\def\tN{{\widetilde N}}
\def\btq{{\tilde{q}}} \def\tbq{{\tilde{q}}} 
\def\tCV{{\widetilde {\cal V}}}

\def\IZ{Z\!\!\!Z}

\def\blank#1{}

%
%
\def\frac#1#2{{\scriptstyle{#1 \over #2}}}                
\def\inv#1{\scriptstyle{1 \over #1}}

\def\ket#1{ | #1 \rangle}
\def\bra#1{ \langle #1 |}
\def\llangle{\langle\!\langle}
\def\rrangle{\rangle\!\rangle}
%
%
\def\CA{{\cal A}}              
       \def\CE{{\cal E}}       
       \def\CH{{\cal H}}       \def\CI{{\cal I}}
              
       \def\CN{{\cal N}}

\def\CV{{\cal V}}

\def\({ \left( }\def\[{ \left[ }
\def\){ \right) }\def\]{ \right] }
%


\def\IR{\relax{\rm I\kern-.18em R}}
\font\cmss=cmss10 \font\cmsss=cmss10 at 7pt
\def\IZ{\relax\ifmmode\mathchoice
{\hbox{\cmss Z\kern-.4em Z}}{\hbox{\cmss Z\kern-.4em Z}}
{\lower.9pt\hbox{\cmsss Z\kern-.4em Z}}
{\lower1.2pt\hbox{\cmsss Z\kern-.4em Z}}\else{\cmss Z\kern-.4em Z}\fi}
\def\inbar{\,\vrule height1.5ex width.4pt depth0pt}
\def\IB{\relax{\rm I\kern-.18em B}}
\def\ID{\relax{\rm I\kern-.18em D}}
\def\IE{\relax{\rm I\kern-.18em E}}
\def\IF{\relax{\rm I\kern-.18em F}}
\def\IG{\relax\hbox{$\inbar\kern-.3em{\rm G}$}}
\def\IH{\relax{\rm I\kern-.18em H}}
\def\II{\relax{\rm I\kern-.18em I}}
\def\IK{\relax{\rm I\kern-.18em K}}
\def\IL{\relax{\rm I\kern-.18em L}}
\def\IM{\relax{\rm I\kern-.18em M}}
\def\IN{\relax{\rm I\kern-.18em N}}
\def\IO{\relax\hbox{$\inbar\kern-.3em{\rm O}$}}
\def\IP{\relax{\rm I\kern-.18em P}}
\def\IQ{\relax\hbox{$\inbar\kern-.3em{\rm Q}$}}
\def\IGa{\relax\hbox{${\rm I}\kern-.18em\Gamma$}}
\def\IPi{\relax\hbox{${\rm I}\kern-.18em\Pi$}}
\def\ITh{\relax\hbox{$\inbar\kern-.3em\Theta$}}
\def\IOm{\relax\hbox{$\inbar\kern-3.00pt\Omega$}}



\def\oh{{1\over 2}}
\def\bz{\bar z}\def\tr{{\rm tr}\,}

\def\Ga{\alpha}


\def\bra{\langle}\def\ket{\rangle}

\font\ttne=cmtt9 
\def\qalg#1{{\ttne q-alg #1}}
\def\hepth#1{{\ttne hep-th #1}}
\def\condmat#1{{\ttne cond-mat #1}}

\def\slh{\widehat{sl}}

\def\tvp{\vrule height 2pt depth 1pt} 
\def\thp{\vrule height 0.4pt width 0.35em}
\def\cc#1{\hfill#1\hfill}
\setbox34=\hbox{$\scriptstyle {p}$} %
\setbox3=\hbox{$\vcenter{\offinterlineskip
\+ \thp&\cr  
\+ \tvp\cc{}&\tvp\cr 
\+ \thp&\cr  
\+ $\!{}^{\vdots}$\cc{}&$\!{}^{\vdots}$\cr 
\+ \thp&\cr  
\+ \tvp\cc{}&\tvp\cr 
\+ \thp&\cr  }$}
\setbox22=\hbox{$\left.\vbox to \ht3{}\right\}$} 
\def\npbox{n_{\copy3\copy22\copy34}}

\def\\#1 {{\tt\char'134#1} }

\def\a{a}\def\b{b}
\def\Omit#1{{}}

\catcode`\@=11
\def\Eqalign#1{\null\,\vcenter{\openup\jot\m@th\ialign{
\strut\hfil$\displaystyle{##}$&$\displaystyle{{}##}$\hfil
&&\qquad\strut\hfil$\displaystyle{##}$&$\displaystyle{{}##}$
\hfil\crcr#1\crcr}}\,}   \catcode`\@=12

\def\encadre#1{\vbox{\hrule\hbox{\vrule\kern1pt\vbox{\kern1pt#1\kern1pt}
\kern1pt\vrule}\hrule}}
\def\encadremath#1{\vbox{\hrule\hbox{\vrule\kern8pt\vbox{\kern8pt
\hbox{$\displaystyle #1$}\kern8pt}
\kern8pt\vrule}\hrule}}

\newdimen\xraise\newcount\nraise
\def\xpoint{\hbox{\vrule height .45pt wNth .45pt}}
\def\udiag#1{\vcenter{\hbox{\hskip.05pt\nraise=0\xraise=0pt
\loop\ifnum\nraise<#1\hskip-.05pt\raise\xraise\xpoint
\advance\nraise by 1\advance\xraise by .4pt\repeat}}}
\def\ddiag#1{\vcenter{\hbox{\hskip.05pt\nraise=0\xraise=0pt
\loop\ifnum\nraise<#1\hskip-.05pt\raise\xraise\xpoint
\advance\nraise by 1\advance\xraise by -.4pt\repeat}}}
\def\bdiamond#1#2#3#4{\raise1pt\hbox{$\scriptstyle#2$}
\,\vcenter{\vbox{\baselineskip12pt
\lineskip1pt\lineskiplimit0pt\hbox{\hskip10pt$\scriptstyle#3$}
\hbox{$\udiag{30}\ddiag{30}$}\vskip-1pt\hbox{$\ddiag{30}\udiag{30}$}
\hbox{\hskip10pt$\scriptstyle#1$}}}\,\raise1pt\hbox{$\scriptstyle#4$}}



\def\IC{\relax\hbox{$\inbar\kern-.3em{\rm C}$}}

\omit{
\def\msy{y }
\message{Do you have the AMS fonts (y/n) ?}\read-1 to \msan
\ifx\msan\msy   
}
\input amssym.def
\input amssym.tex
\def\IZ{\Bbb Z}\def\IR{\Bbb R}\def\IC{\Bbb C}\def\IN{\Bbb N}
\def\II{\Bbb I}\def\IP{\Bbb P}
\def\gA{\goth A}


\def\irrep{irreducible representation}
\def\ib{\bar i}

\newdimen\xraise \newcount\nraise
\def\xpoint{\hbox{\vrule height .45pt width .45pt}}
\def\udiag#1{\vcenter{\hbox{\hskip.05pt\nraise=0\xraise=0pt
\loop\ifnum\nraise<#1\hskip-.05pt\raise\xraise\xpoint
\advance\nraise by 1\advance\xraise by .4pt\repeat}}}
\def\ddiag#1{\vcenter{\hbox{\hskip.05pt\nraise=0\xraise=0pt
\loop\ifnum\nraise<#1\hskip-.05pt\raise\xraise\xpoint
\advance\nraise by 1\advance\xraise by -.4pt\repeat}}}
\def\bdiamond#1#2#3#4{\raise1pt\hbox{$\scriptstyle#2$}
\,\vcenter{\vbox{\baselineskip12pt
\lineskip1pt\lineskiplimit0pt\hbox{\hskip10pt$\scriptstyle#3$}
\hbox{$\udiag{30}\ddiag{30}$}\vskip-1pt\hbox{$\ddiag{30}\udiag{30}$}
\hbox{\hskip10pt$\scriptstyle#1$}}}\,\raise1pt\hbox{$\scriptstyle#4$}}
\def\badiamond#1#2#3#4{\raise1pt\hbox{$\scriptstyle#2$}
\,\vcenter{\vbox{\baselineskip12pt
\lineskip1pt\lineskiplimit0pt\hbox{\hskip10pt$\scriptstyle#3$}
\hbox{$\swarrow\!\searrow$}\vskip4pt
\hbox{$\searrow\swarrow$}
\hbox{\hskip10pt$\scriptstyle#1$}}}\,\raise1pt\hbox{$\scriptstyle#4$}}


\font\ttne=cmtt9 
\font\itne=cmti9 
\font\slne=cmsl9 
\font\bfne=cmbx9 
	{$\left.\right.$}

\lref\KFF{V.G. Kac, {\itne Lect. Notes in Phys.} {\bfne 94}
(1979) 441-445\semi  B.L. Feigin and D.B.  Fuchs, 
{\itne Funct. Anal. and Appl.} {\bfne 16} (1982) 114-126; {\itne ibid.} 
{\bfne 17} (1983) 241-242.  }

\lref\Ka{V. Kac, {\itne Infinite dimensional algebras}, 
Cambridge University P.; 
V.G. Kac and D.H. Peterson, {\itne Adv. Math.} {\bfne 53} (1984) 125-264\semi
 J. Fuchs, {\itne Affine Lie Algebras and Quantum Groups{}},
Cambridge Univ. Pr. 1992.  }

\lref\GKO{P. Goddard, A. Kent and D. Olive, {\itne Comm. Math. Phys.}
{\bfne 103} (1986) 105-119.}

\lref\ZF{A.B. Zamolodchikov and V.A. Fateev, 
{\itne Sov. J. Nucl.Phys.} {\bf 43} (1986) 657-664.}  
 
\lref\DF{
 Vl.S. Dotsenko and  V.A. Fateev, {\itne Nucl. Phys.} {\bfne B 251}
[FS13] (1985) 691-734.}

\lref\DFMS{P. Di Francesco, P. Mathieu and D. S\'en\'echal, {\itne Conformal 
Field Theory}, Springer Verlag 1997.}

\lref\EV{E. Verlinde, {\itne  Nucl. Phys.} {\bfne B300} [FS22]
(1988) 360-376. } 

\lref\Camod{ 
J. Cardy, {\itne Nucl Phys} {\bfne B 270} (1986) 186-204.}

\lref\Sono{H. Sonoda, {\itne Nucl. Phys.} {\bfne B281} (1987) 546-572;
{\bfne B 284} (1987) 157-192.}

\lref\MS{G. Moore and N. Seiberg,  
{\itne Comm. Math. Phys.} {\bfne 123} (1989) 177-254;
{\itne Lectures on RCFT, Physics, Geometry and Topology}, Plenum Press,
New York, 1990.}

\lref\CIZK{A. Cappelli, C. Itzykson and J.-B. Zuber,
{\itne Nucl. Phys.} {\bfne B280} [FS18] (1987)
445-465; {\itne Comm. Math. Phys.} {\bfne 113} (1987) 1-26
\semi A. Kato, {\itne Mod. Phys. Lett.} {\bfne A2} (1987) 585-600.}

\lref\Ga{T. Gannon, {\itne Comm. Math. Phys.} {\bfne 161} (1994) 233-263;
{\itne The Classification of $SU(3)$ Modular Invariants Revisited},
\hepth{9404185}.}

\lref\IDG{C. Itzykson, {\itne Nucl. Phys.} (Proc. Suppl.) {\bfne
5B} (1988) 150-165 
\semi P. Degiovanni, {\itne Comm. Math. Phys.} {\bfne 127} (1990) 71-99. }

\lref\Gannew{ 
T. Gannon, {\itne The monstruous moonshine and the classification of CFT}, 
\hepth{9906167}. }

\lref\Zbarilo{J.-B. Zuber, {\itne CFT, BCFT, ADE and all that}, 
to appear in the proceedings of the Bariloche Summer School, 
``Quantum Symmetries in Theoretical Physics and Mathematics'', 
Jan 2000, eds R. Coquereaux, A. Garcia and 
R. Trinchero,  \hepth{0006151}.}

\lref\BPZ{R.E. Behrend, P.A. Pearce and J.-B. Zuber, 
{\itne J.Phys.} {\bfne A31} (1998) L763-L770, \hepth{9807142}.}

\lref\BPPZ{R.E. Behrend, P.A. Pearce, V.B. Petkova and J.-B. Zuber, 
{\itne Nucl. Phys.} {\bfne B 579} (2000) 707-773, \hepth{9908036}.}

\lref\AOS{I. Affleck, M.  Oshikawa and H. Saleur, {\itne J. Phys A}
{\bfne 31} (1998) {5827-5842}, \condmat{9804117}. }

\lref\CaL{J.L. Cardy and D.C. Lewellen, {\itne Phys. Lett. B}
{\bfne 259} (1991) 
{274-278}
\semi D.C. Lewellen, {\itne Nucl. Phys.} {\bfne B 372} (1992) {654-682}.}

\lref\Schbud{C. Schweigert, lectures at  the Summer  School
{\slne Nonperturbative Quantum Field Theoretic Methods and their
Applications},  
19-21  August 2000, E\"otv\"os University, Budapest, Hungary, to appear.}

\lref\HSal{H. Saleur, 
{\slne Lectures on Non Perturbative Field Theory and Quantum
Impurity Problems, I and II}, \condmat{9812110}, \condmat{0007309}.}

\lref\Dorey{P. Dorey, lectures at this school.}

\lref\Wa{G. Watts, private communication.}
\lref\Watts{
K. Graham, I. Runkel and G.M.T. Watts, 
{}PRHEP-tmr2000/040  (Proceedings of the TMR network conference
{\itne Nonperturbative Quantum Effects 2000}),
\hepth{0010082}.}

\lref\Affleck{L. Chim, {\itne Int. J. Mod. Phys.} {\bfne A11}
(1996) 4491-4512,  
\hepth{9510008}\semi
I. Affleck,
{\slne Edge critical behaviour of the 2-dimensional tri-critical Ising model}, 
\condmat{0005286}.}

\lref\Reck{A. Recknagel, D. Roggenkamp, V. Schomerus,
{\slne On relevant boundary perturbations of unitary minimal models},
\hepth{0003110}.}

\lref\Ru{I. Runkel, {\itne Nucl. Phys.} {\bfne B 549} (1999)
{563-578}, 
\hepth{9811178}.} 

\lref\Rudeu{I. Runkel, {\itne Nucl. Phys.} {\bfne B 579} (2000)
{561-589},  \hepth{9908046}.}

\lref\GHJ{F.M. Goodman, P. de la Harpe and V.F.R. Jones, {\itne Coxeter
Graphs and Towers of Algebras}, Springer-Verlag, Berlin (1989).}

\lref\Jo{V.F.R. Jones, {\itne Invent. Math.} {\bfne 72} (1983) 1-25.}

\lref\JW{M. Jimbo, T.  Miwa and M. Okado, {\itne  Lett. Math. Phys.}
{\bfne 14} (1987) 123-131 ;  {\itne Comm. Math. Phys.} {\bfne 119} (1988) 
543-565
\semi
H. Wenzl,  {\itne Inv. Math.} {\bfne 92} (1988) 349.}   

\lref\WDVV{
T. Eguchi and S.K. Yang, {\itne Mod. Phys. Lett.} {\bfne A5} (1990) 1693-1701
\semi
R. Dijkgraaf, E. Verlinde and H. Verlinde, Nucl. Phys. 
{\bfne B352} 59-86 (1991):
in {\itne String Theory and Quantum Gravity},
proceedings of the 11990 Trieste Spring School, M. Green et al. {\itne eds.},
World Sc. 1991.}

\lref\VPun{V. Pasquier, {\itne Nucl. Phys.} {\bfne B285} [FS19]
(1987) 162-172.} 

\lref\VPdeu{V. Pasquier, {\itne J. Phys.} {\bfne A20} (1987) 5707-5717.}

\lref\LVWM{
 W. Lerche, C. Vafa and N. Warner,
{\itne Nucl. Phys. } {\bfne B 324} (1989) 427-474
\semi E. Martinec, in 
{\itne Criticality, 
catastrophes and compactifications}, in 
{\itne Physics and mathematics of strings}, V.G. Knizhnik memorial volume,
L. Brink, D. Friedan and A.M. Polyakov eds., World Scientific 1990
\semi
P. Howe and P. West, {\itne Phys. Lett.} {\bfne B223} (1989) 377-385; 
{\itne ibid.} {\bfne B227}  (1989) 397-405.
}

\lref\Zubkyoto{J.-B. Zuber, {\slne C-algebras and their applications to
reflection groups and conformal field theories}, lectures at RIMS, 
Kyoto December 1996, \hepth{9707034}. }

\lref\OcnEK{A. Ocneanu, ``Quantized groups, string algebras and Galois theory
for algebras'', in {\itne Operator Algebras and Applications},  D. Evans and
Takesaki eds, 1988, pp. 119-172
\semi D. Evans and Y. Kawahigashi,
{\itne Publ. RIMS, Kyoto Univers. } {\bfne 30} (1994) 151-166.}

\lref\Pasq{V. Pasquier, {\slne Mod\`eles Exacts Invariants Conformes}, 
Th\`ese d'Etat, Orsay, 1988.}

\lref\DFZun{P. Di Francesco and J.-B. Zuber,
{\itne Nucl. Phys.} {\bfne B338} (1990) 602-646
.}

\lref\Soch{N. Sochen, {\itne Nucl. Phys.} {\bfne B360} (1991) 613-640. }

\lref\PZun{V.B. Petkova and J.-B. Zuber, {\itne Nucl. Phys.} {\bfne B 438 } 
(1995) 347-372, \hepth{9410209}. }

\lref\PZdeu{V.B. Petkova and J.-B. Zuber, {\itne Nucl. Phys.} {\bfne B 463} 
(1996) 161-193, \hepth{9510175}; {\slne Conformal Field Theory and 
Graphs} \hepth{9701103}. }

\lref\PZtw{V.B. Petkova and J.-B. Zuber,
{\slne Generalised twisted partition functions},
{\itne Phys. Lett. B } to appear,  \hepth{0011021}.}

\lref\Ocn{A. Ocneanu, {\itne Paths on Coxeter Diagrams},
in {\itne Lectures on Operator Theory}
Fields Institute Monographies, Rajarama Bhat et al edrs, 
AMS 1999.}

\lref\BE{ 
{J. B\"ockenhauer  and D.E. Evans, } 
{\itne Comm.Math. Phys.} 
{\bfne 197} (1998) 361-386; {\itne ibidem} {\bfne 200} 
(1999) {57-103}, \hepth{9805023}; {\itne ibidem} {\bfne 205} (1999) 183-228,  
\hepth{9812110}.}

\lref\BEK{ J. B\"ockenhauer, D.E. Evans, and Y. Kawahigashi, 
{\itne Comm.Math. Phys.} {\bfne 208} (1999) 429-487, {\ttne math-OA 9904109};
{\itne ibidem} {\bfne 210} (2000) 733-784.} 

\lref\Xu{F. Xu, {\itne Comm. Math. Phys.} {\bfne 192} (1998) 349-403. }

\lref\AO{A. Ocneanu, lectures at the school
``Quantum Symmetries in Theoretical Physics and Mathematics'', 
January 2000, Bariloche, R. Coquereaux, A. Garcia and 
R. Trinchero eds, to appear.}

\lref\Ztani{J.-B. Zuber, {\itne Generalized Dynkin diagrams and root systems
and their folding}, Proccedings of the Taniguchi meeting, Kyoto Dec 1996, 
{\itne Topological Field Theory, Primitive Forms and Related Topics},
M. Kashiwara, A. Matsuo, K. Saito and I. Satake edrs, Birkh\"auser.}

\lref\GZV{S.M. Gusein-Zaide and A.N. Varchenko, {\itne Verlinde algebras 
and the intersection form on vanishing cycles}, \hepth{9610058}.} 

\lref\RS{
A. Recknagel and V. Schomerus, {\itne Nucl. Phys.} {\bfne B 531}
{(1998)} {185-225}
.}

\lref\FS{J. Fuchs and C. Schweigert, {\itne Nucl. Phys.} {\bfne B
530 } (1998) 99-136, \hepth{9712257}.} 

\lref\Cabc{J.L. Cardy, {\itne Nucl. Phys.} {\bfne 324} (1989) {581-596}.}

\lref\PSS{{G. Pradisi, A. Sagnotti and Ya.S. Stanev}, 
{\itne Phys. Lett.} {\bfne 381} {(1996)} {97-104}.}

\lref\PaSa{V. Pasquier and H. Saleur {\itne Nucl. Phys. } {\bfne B 330} (1990)
523-556.}

\lref\Kos{B. Kostant, {\itne Proc. Natl. Acad. Sci. USA} {\bfne 81} (1984) 
5275-5277;  
{{\itne Ast\'erisque (Soci\'et\'e Math\'ematique de France)},}{
}(1988) {209-255, } 
{\itne The McKay Correspondence, the Coxeter Element and Representation
Theory}.}

\lref\Dor{
P. Dorey, {\itne Int. J. Mod. Phys.}{\bfne 8} (1993) 193-208.}

\lref\BCDS{H. Braden, E. Corrigan, P.E. Dorey  and R. Sasaki,
{\itne Nucl. Phys.} 
{\bfne B 338} (1990) {689-746}. }

\lref\MCOBSS{
B.M. McCoy and W. Orrick {\itne Phys. Lett. }{\bfne A 230}
(1997) {24-32} 
\semi
{M.T. Batchelor and K.A. Seaton, K.A., }{\tt cond-mat 9803206},  
{\slne Excitations in the
diluate $A_L$ lattice model: $E_6$, $E_7$ and $E_8$ mass spectra}
\semi
{J. Suzuki}, {\tt cond-mat 9805241}, {\slne Quantum Jacobi-Trudi Formula and
$E_8$ Structure in the Ising Model in a Field}.}

\lref\BI{E. Bannai, T. Ito, {\itne Algebraic Combinatorics I: Association
Schemes}, Benjamin/Cummings (1984).}

\lref\DFZdeu{P. Di Francesco and J.-B. Zuber,
in {\itne Recent Developments in Conformal Field Theories}, Trieste Conference
1989, S. Randjbar-Daemi, E. Sezgin and J.-B. Zuber eds., World Scientific
1990 \semi
P. Di Francesco, {\itne Int. J. Mod. Phys.} {\bfne A7} (1992) 407-500.}

\lref\BP{R.E. Behrend and P.A. Pearce, {\itne J. Phys. A} {\bfne 29} (1996) 
7827-7835; {\itne Int. J. Mod. Phys.} {\bfne 11} (1997) 2833-2847.} 

\lref\BPde{R.E. Behrend and P.A. Pearce, 
{\slne Integrable and Conformal Boundary Conditions for $\slh(2)$
$A-D-E$ Lattice Models and Unitary Conformal Field Theories}, 
\hepth{0006094}, {\itne J. Stat Phys.} to
appear. }

\lref\PZtmr{V.B. Petkova and J.-B. Zuber, 
{}PRHEP-tmr2000/038  (Proceedings of the TMR network conference
{\itne Nonperturbative Quantum Effects 2000}),
 \hepth{0009219}.}

\lref\PZnew{V.B. Petkova and J.-B. Zuber, {\slne The many faces of 
Ocneanu cells}, \hepth{0101151}.}

\lref\BSz{ G. B\"ohm and K. Szlach\'anyi, 
{\itne Lett. Math. Phys.} {\bfne 200} {(1996)} {437-456},
\qalg{9509008}
\semi G. B\"ohm,  {\slne Weak $C^*$-Hopf Algebras and their
Application to Spin Models}, PhD Thesis, Budapest 1997.}



%
\font\tenbf=cmbx10
\font\tenrm=cmr10
\font\tenit=cmti10
\font\ninebf=cmbx9
\font\ninerm=cmr9
\font\nineit=cmti9
\font\eightbf=cmbx8
\font\eightrm=cmr8
\font\eightit=cmti8
\font\sevenrm=cmr7 	%

\def\sectiontitle#1\par{\vskip0pt plus.1\vsize\penalty-250
 \vskip0pt plus-.1\vsize\bigskip\vskip\parskip
 \message{#1}\leftline{\tenbf#1}\nobreak\vglue 5pt}
\hsize=4.7truein
\vsize=7truein
\parindent=15pt
\nopagenumbers
\baselineskip=13pt
\pageno=1  

\line{\eightrm
Proceedings of Budapest School and Conference
\hfil}
\line{\eightrm  Nonperturbative Quantum Field Theoretic Methods 
 and their Applications,  August 2000, 
\hfil}
\line{\eightrm $\copyright$ World Scientific Publishing Company\hfil}

\vglue 5pc
\baselineskip=13pt
\headline{\ifnum\pageno=1\hfil\else
{\ifodd\pageno\rightheadline \else \leftheadline\fi}\fi}
\def\rightheadline{\hfil\eightit
Conformal Boundary Conditions
\quad\eightrm\folio}
\def\leftheadline{\eightrm\folio\quad
\eightit
\omitt{V.B. Petkova and } 
J.-B. Zuber
\hfil}
\voffset=2\baselineskip
\centerline{\tenbf
CONFORMAL \hskip 0.1cm BOUNDARY\hskip 0.1cm CONDITIONS }
\centerline{\tenbf 
and \hskip 0.1cm what \hskip 0.1cm they \hskip 0.1cm teach \hskip 0.1cm us}
\vglue 24pt
\omitt{\centerline{\eightrm
VALENTINA B. PETKOVA}
\omit{
\baselineskip=12pt
\centerline{\eightit
School of Computing and Mathematics}
\baselineskip=10pt
\centerline{\eightit
University of Northumbria}
\baselineskip=10pt
\centerline{\eightit
NE1 8ST Newcastle upon Tyne, UK}}
\baselineskip=12pt
\centerline{\eightit
Institute for Nuclear Research and Nuclear Energy }
\baselineskip=10pt
\centerline{\eightit
72 Tzarigradsko Chaussee,  1784 Sofia, Bulgaria,}
\baselineskip=12pt
\centerline{\eightit
School of Computing and Mathematics}
\baselineskip=10pt
\centerline{\eightit
University of Northumbria}
\baselineskip=10pt
\centerline{\eightit
NE1 8ST Newcastle upon Tyne, UK}
\centerline{\eightit
E-mail: valentina.petkova@unn.ac.uk }
\medskip
\centerline{and}}
\medskip
\centerline{\eightrm
JEAN-BERNARD ZUBER
}
\baselineskip=12pt
\centerline{\eightit
Service de Physique Th\'eorique 
}
\baselineskip=10pt
\centerline{\eightit
CEA-Saclay}
\baselineskip=12pt
\centerline{\eightit
F-91191 Gif-sur-Yvette,  France }
\centerline{\eightit
E-mail: zuber@spht.saclay.cea.fr}

\vglue 20pt
\centerline{  }

{\rightskip=1.5pc
\leftskip=1.5pc
\eightrm\parindent=1pc \baselineskip=12pt
\noindent 
The question of boundary conditions in conformal
field theories is discussed, in the light of recent progress. 
Two kinds of boundary conditions are examined, along open 
boundaries of the system, or along closed curves or ``seams''. 
Solving  consistency
conditions known as Cardy equation is shown to amount to the 
algebraic problem of finding integer valued representations of 
(one or two copies of)
the fusion algebra. Graphs encode these boundary conditions 
in a natural way, but are also relevant in several aspects of 
physics ``in the bulk''. Quantum algebras attached to these 
graphs contain information on structure constants of the operator
algebra, on the Boltzmann weights of  the corresponding integrable 
lattice models etc.
Thus the study of boundary conditions in Conformal Field
Theory offers a new perspective on several old physical problems 
and offers an explicit realisation of recent mathematical concepts. 

\vglue4pt
\leftskip=2.5pc\rightskip=2.5pc
\parindent=2.5pc \baselineskip=12pt
}
\baselineskip=13pt
\overfullrule=0pt
%
\vskip 0.5cm

%


\secno=-1
\newsec{Introduction}
\nind
The study of boundary conditions in conformal field theories (CFT)
and in the related integrable models
has been experiencing a renewal of interest over the last three or four
years. This has been caused by its relevance in string and brane 
theory on the one hand, and in various  problems of condensed matter
on the other: see the lectures of C. Schweigert at this school
for an introduction and references to the first subject, 
and \HSal\ 
for the second. As a result, 
there has been a blossoming of papers studying the possible 
boundary conditions, the boundary fields  and their  couplings 
in the framework of CFT (see \BPPZ\ for a fairly extensive 
bibliography as of mid 99); a systematic discussion of boundary 
conditions preserving the integrability, both in lattice models \BP\
and in (classical or quantum) field theories \Dorey; and an investigation 
of what happens to a critical system in the presence of boundary 
perturbations, its renormalisation group flows, etc: 
see in particular the lectures by G. Watts at this school and  \Watts.  
At the same time, new and unexpected connections with ``pure'' mathematics
--operator algebras, quantum symmetries-- have also been revealed.

The purpose of these lectures is of course {\it not} to present exhaustively
all these interesting developments, but just to offer a pedagogical
(and maybe somewhat biaised) introduction to their simplest aspects
and to some of the recent progress. 
After briefly recalling basic facts on CFT, their chiral
constituents and how they are assembled into physically sensible
theories, we turn to the discussion of boundary conditions.  
We show how solving the consistency
condition known as Cardy equation amounts to the
algebraic problem of finding non negative integer valued matrix 
representations of the fusion algebra. These matrices are the adjacency 
matrices of graphs, which thus encode the boundary conditions (b.c.)
in  a natural way (sect.~2). The study of 
the operator algebra of boundary fields (sect.~3)
and of possible twisted b.c. (sect.~4)
exposes new algebraic features attached to these graphs
 (sect.~5).  The latter also  contain information 
on Boltzmann weights of associated lattice integrable models, as
we mention briefly in section 6. 

\newsec{A lightning review of CFT}
\nind
This section is devoted to a fast summary of  concepts and notations
in rational conformal field theories (RCFT).

\subsec{Chiral data of RCFT}
\nind A rational conformal field theory is defined in terms of a certain number 
of data. The first set of data specifies the properties of each chiral half, 
i.e. of the holomorphic or of the antiholomorphic sector of the theory. 
One is 
given a certain {\it chiral algebra}, $\gA$: it may be the Virasoro 
algebra Vir itself, with its generators $L_n$, $n\in \IZ$,  or 
equivalently the energy-momentum tensor
$T(z)=\sum_{n\in\IZ} z^{-n-2} L_n$. It may also be one of the 
extensions of Vir: 
superconformal algebra, current 
algebra, $W$-algebra etc. One is also given a {\it finite} set $\CI$
of irreducible representation spaces (modules) 
$\{\CV_i\}_{i\in \CI}$ of $\gA$.  Each of 
these representations of $\gA$ is also a  representation (reducible
or irreducible) of Vir, with a central charge $c$ and with a
conformal weight
(the lowest eigenvalue of $L_0$) denoted $h_i$. Let's recall for future 
use that $c$ also specifies the coefficient of the anomalous
term in the transformation of the energy-momentum tensor $T(z)$ under 
an analytic change of coordinate $z \mapsto \zeta(z)$
\eqn\varT
{ {\tilde T}(\zeta)=\Big({\partial z\over \partial\zeta} \Big)^2
 T(z) +{c\over 12} \{z,\zeta\} }
where $\{z,\zeta\} $ denotes the schwarzian derivative
\eqn\schwarz
{ \{z,\zeta\}=
{{\partial^3 z\over\partial \zeta^3}\over
{\partial z\over\partial \zeta}} 
-{3\over 2} \Bigg({{\partial^2 z\over\partial \zeta^2}\over
{\partial z\over\partial \zeta}} \Bigg)^2 \ .}
By convention, the label $i=1$ denotes the identity representation (for which
$h_1=0$).  Finally, we denote 
by $\CV_{i^*}$ the complex conjugate representation of $\CV_i$; 
 the identity representation is self-conjugate, $1^*=1$.

Each of these representations is graded for the action of the Virasoro
generator $L_0$ : 
all the  eigenvalues of $L_0$ differ from the lowest one, $h_i$, 
 by a non negative 
integer
\foot{This is strictly true only for algebras, whose generators are 
integrally graded. In a superconformal algebra, or in a parafermionic 
algebra, the grading would be fractional.},
and the eigenspace of eigenvalue $h_i+p$ has a certain dimension $\#_p^{(i)}$.
It is  natural to introduce the ``character'' of the representation
$\CV_i$, which is, up to a prefactor, the generating function of these
dimensions
\eqn\chara
{\chi_i(q)=\tr q^{L_0-{c\over 24}}=q^{h_i-{c\over 24}}
\sum_{p=0}^\infty \, \#_p^{(i)}\,q^p\ .}

{\petit \baselineskip=12pt
The simplest example is given by the integrable 
representations of the 
affine (current) algebra $\slh(2)$. For an integer value of the central
charge of the affine algebra (or level) $k$, the Virasoro
central charge is $c=3k/(k+2)$, and  there is a finite set of
integrable representations, labelled by an integer $j$: $1\le j\le k+1$, 
whose conformal weights are $h_j=(j^2-1)/4(k+2)$. In
that case, the conjugation is trivial: $\CV_j=\CV_{j^*}$. 
For  the representation $(j, k)$, the character reads
\eqn\sldch{\chi_j(q) ={1\over \eta^3(q) }\sum_{p=-\infty}^\infty
(2(k+2)p+j)\,q^{{(2(k+2)p+j)^2\over 4(k+2)}}\ ,}
where $\eta(q)=q^{{1\over 24}}\prod_1^\infty (1-q^n)$
 is the Dedekind eta function. Such characters are called ``specialized 
characters'' since they count states according to their $L_0$ grading only.
Non-specialized characters can be introduced, which are sensitive to
the Cartan algebra generator ${J}_0$
\eqn\spech{ \chi_j(q,{u})= \tr\, q^{L_0-{c\over 24}} e^{2\pi i {u}{J}_0}
\ .}
The expressions of non-specialized characters and/or for higher rank 
algebras may be found in \Ka.

Another example is provided by the minimal $c<1$ theories. They are
parametrized by a pair of coprime integers $p$ and $p'$, and the 
central charge takes the values
$c=1-6(p-p')^2/pp'$. The irreducible representations of Vir are
labelled by a pair of integers $(r,s)$, $1\le r\le p'-1$, 
$1\le s\le p-1$, modulo the identification $(r,s)\equiv (p'-r,p-s)$. Their
conformal weights read
\eqn\hmin{h_{(r,s)}=h_{(p'-r,p-s)}={(rp-sp')^2-(p-p')^2\over 4 pp'}\,. }
Again the conjugation acts trivially. 
The character of this \irrep\ reads, with the notations $\lambda:=(rp-sp')$,
$\lambda':=(rp+sp')$
\eqn\chirrep
{\chi_{(r,s)}(q)={1\over \eta(q)}
\sum_{n\in \IZ} \( q^{(2npp'+\lambda)^2\over 4pp'}-
q^{(2npp'+\lambda')^2\over 4pp'}\) \ . }
}

{\it Modular transformations}\nind
In the previous expressions, $q$ is a dummy variable.
If, however, $q$ is regarded as a complex variable of modulus less than one
and written as $q=\exp 2i\pi \tau$, with $\tau$ a complex number of positive 
imaginary part, one may prove that the sum converges and has remarkable 
modular properties. Under a $PSL(2,\IZ)$ transformation of $\tau$:
$$ \tau\mapsto {a+b\tau \over c+ d \tau}\,, \qquad a,b,c,d\in \IZ,\ ad-bc=1\ ,
$$
the set of functions $\chi_i$ transforms linearly, and in fact supports 
a  unitary representation of (the  double cover of) $PSL(2,\IZ)$.
 In particular
there exists a unitary matrix $S$ implementing the transformation 
$\tau \mapsto -1/\tau$. If $\tilde q:= \exp -{2i\pi\over \tau}$, there exists
a unitary $|\CI|\times |\CI|$ matrix $S$ such that
\eqn\Strans
{\chi_i( q) =\sum_{j\in \CI} S_{ij} \chi_j(\tilde q)\ . }
Moreover the matrix $S$ satisfies
$S^T=S$,
\quad $(S_{ij})^*=S_{i^*j}
=S_{ij^*}$, $S^2=C={\rm the\ conjugation\ matrix}$
defined by $C_{ij}=\delta_{ij^*}$, \quad $S^4=I$.
\foot{The fact that $S^2=C$ rather than $S^2=I$
as expected from the transformation $\tau \mapsto -1/\tau$ signals that we
are dealing with a representation of a double 
covering of the modular group.}

\nind
{\petit \baselineskip=12pt
 For the $c<1$ minimal representations,
$$\CI=\{(r,s)\equiv (p'-r, p-s)\ ;\quad  1\le r\le p'-1,\ 1\le s\le p-1\}$$
and the $S$ matrix reads
\eqn\Smin
{S_{(r,s),(r's')}=\sqrt{{8\over pp'}}(-1)^{(r+s)(r'+s')}
\sin\pi rr'{p-p'\over p'}\,\sin\pi ss'{p-p'\over p} }
\nind 
For the $\slh(2)$ affine algebra, at level $k$, 
 for which $ \CI=\{1,2,\cdots, k+1\}$, one finds
\eqn\Ssld{S_{jj'} = \sqrt{2\over k+2} \sin{\pi jj'\over k+2}\ ,\quad
j,j'\in \CI\,\ .}
The expression for more general affine algebras may be found in \Ka.
For non-specialised characters, the transformation reads:
\eqn\trunsp{\chi_i(q,{u})=
e^{-i k\pi {u}^2 /\tau} \,
 \sum_{j\in  \CI} S_{ij} \chi_j(\tilde q, -{u}/\tau) \ .}
}

\nind{\petit 
 One notes that the $S$ matrix of the minimal case \Smin\
is ``almost'' the tensor product of two matrices of the
form \Ssld, at two different  levels $k=p-2$
and $k'=p'-2$. This would be true for $|p-p'|=1$ and 
if one could omit the
identification $(r,s)\equiv (p'-r, p-s)$. This is of course not a
coincidence but reflects the ``coset construction'' of $c<1$
representations of Vir out of the affine algebra $\slh(2)$ \GKO. 
We shall encounter below again this fact that minimal cases 
are ``almost'' the tensor products of the $\slh(2)$ ones. }

\medskip

{\it Fusion Algebra}
\nind
Another concept of crucial importance for our discussion is
that of fusion algebra. Fusion is an associative and commutative
operation among
representations of chiral algebras of RCFTs, inherited from the operator
product algebra of Quantum Field Theory. 
It looks similar to the usual tensor product of
representations, but contrary to the latter, it is consistent 
with the finiteness of the set $\CI$ and it preserves the
central elements (instead of adding them). We shall refer to the
literature \refs{\DFMS} for a systematic discussion of this concept,
and just introduce a notation $\star$ to denote it and  distinguish
it from the tensor product. It is natural to decompose the
fusion of two representations of a chiral algebra on the irreducible 
representations, thus defining  ``fusion multiplicities''
\eqn\fuscoef
{  \CV_i \star \CV_j =\oplus_k\, \CN_{ij}{}^k\, \CV_k,
\qquad  \CN_{ij}{}^k \in \IN\ .}

There is a remarkable formula, due to Verlinde \EV, expressing these
multiplicities in terms of the unitary  matrix $S$:
\eqn\verl
{ \CN_{ij}{}^k=\sum_{\ell\in\CI}{S_{i \ell}\, S_{j\ell}
\left(S_{k\ell}\right)\!{}^*
\over S_{1\ell}}\ .}
\nind{\petit We will restrict to RCFT for which the Verlinde
formula \verl\ produces non-negative integers; this excludes e.g., the
fractional level admissible representations of affine algebras, 
for which there is
no such direct relation between the fusion and modular properties.} 

Note that, regarded as matrices $\CN_i=\{\CN_{ij}{}^k\}$, the $\CN$
form a representation (the regular representation) of the fusion 
algebra
\eqn\regfa{\CN_i\,\CN_j=\sum_{k\in \CI}\,\CN_{ij}{}^k\, \CN_k\ .}


{\it Chiral Vertex Operators}\nind
The field theoretic description of chiral halves of CFT (or ``chiral 
cft'') makes use of ``chiral vertex operators'' (CVO): $\phi_{ij}^{\,k}(z)$
is a $z$-dependent homomorphism  from $\CV_i\star\CV_j$ to 
$\CV_k$ and may be diagrammatically depicted as 
\hbox{\raise -2mm\hbox{\epsfxsize=15mm\epsfbox{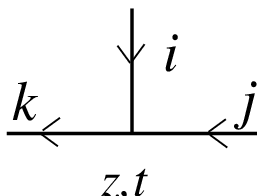}}}.
In fact, there are as many as
$\CN_{ij}{}^k$ such independent intertwiners and the notation 
$\phi_{ij;t}^{\,k}(z)$ and the diagram have to be supplemented 
by a multiplicity label $t\in \{1, 2,\cdots,\CN_{ij}{}^k \}$. 
CVO may be composed (``fused'') as 
$\phi_{il}^{\,m}(z_1)\phi_{jk}^{\,l }(z_2)$ and there is 
an invertible matrix 
$F_{lp}\left[i\atop m\right.\!\!\left.j\atop k\right]^{u\,t}_{t_1\, t_2}$
expressing that 
there are two distinct but equivalent ways of intertwining 
$ \CV_i \star \CV_j \star \CV_k$ and $\CV_m\,$:\ 
\hbox{\raise -2mm\hbox{\epsfxsize=5cm\epsfbox{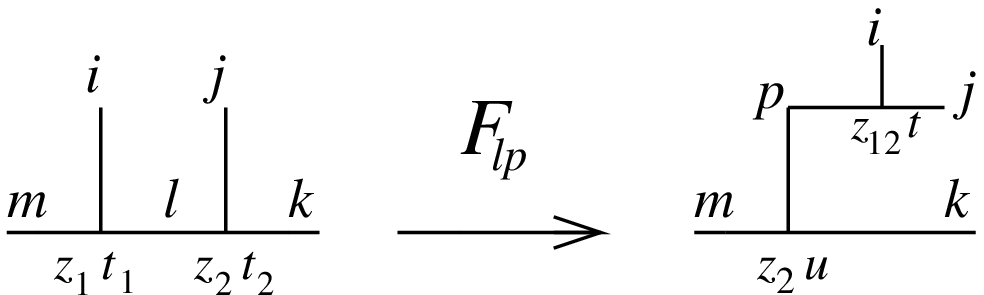}}}.
Note that eq. \regfa\ is a necessary condition of existence of this 
change of basis, since it expresses the equality of the dimensions 
of the two sides: $\sum_l \CN_{jk}{}^l\CN_{il}{}^m=\sum_p 
\CN_{ij}{}^p \CN_{pk}{}^m $.  The matrix $F$ satisfies 
a quintic (``pentagon'') identity expressing the consistency 
(the associativity) in the fusion of three CVOs.  There is also the 
operation of braiding and the matrix
$B_{lp}\left[i\atop m\right.\!\!\left.j\atop k\right](\pm)$
which relates  $\phi_{il}^{\,m}(z_1)\phi_{jk}^{\,l }(z_2)$
and  $\phi_{jp}^{\,m}(z_2)\phi_{ik}^{\,p }(z_1)\,$, 
see \MS\ for a systematic discussion. 

Notice finally that the matrix $S$ admits an extension, 
denoted $S(j)$, such that
$S(1)=S$. The matrix $S(p)$ gives the modular transformation of
1-point conformal block $\bra \phi_{pi}^i\ket_{\tau}$ on the torus. 
See \MS\ for the explicit expression of $S(p)$ in the simplest cases.

\medskip

The data ${c}$  (or $k$, etc), $\{\CV_i, h_i\}_{i\in\CI}$, $S(p)_{ij}$,
${\CN}_{ij}{}^k$, 
$F\left[i\atop k\right.\!\!\left. j\atop l\right]$
and  $B\left[i\atop k\right.\!\!\left. j\atop l\right](\pm)$ 
form what may be  called  the ``chiral data'' of the RCFT.
They are the basic ingredients in the construction of physical theories.
The consistency of the latter imposes however that 
 adequate conditions 
be satisfied by their spectral data and by the structure constants of their
Operator Product Algebra (OPA). This is what we shall consider now, examining 
in turn the cases of the theory ``in the bulk'', i.e. in the absence 
of any boundary, and then in the half-plane.


\subsec{Spectral and OPA  Data in the Bulk}

\nind
In the plane punctured at the origin, equipped with the coordinate $z$,
or equivalently on the cylinder of perimeter $l$ with the coordinate $w$,
with the conformal mapping from the latter to the former $z=\exp -2\pi i w/l$,
a given RCFT is described by a Hilbert space $\CH_P$. This Hilbert space is
decomposable into a {\it finite} sum of \irrep s of {\bf two} copies
of the chiral algebra (Vir or else), associated with the holomorphic and
anti-holomorphic sectors of the theory:
\eqn\hilbert{
\CH_P=\oplus \N_{j\bar j} \CV_j\otimes \overline{\CV_{\bar j}}\ , }
with (non negative integer) multiplicities $\N_{j\bar j}$.
By the state-field correspondence, \hilbert\ also describes
the spectrum of {\it primary fields} of the theory, i.e. of those
fields that transform as heighest weight representations of 
$\gA\otimes\gA$.

A convenient way to encode the information \hilbert\ is to look at
the partition function of the theory on a torus ${\Bbb T}$. Up to a global
dilatation, irrelevant here, a torus may be defined
by its modular parameter $\tau$, $\Im m\,\tau>0$, 
such that its two periods are $1$ and $\tau$. Equivalently, 
it may be regarded as the quotient of the complex plane by the lattice
generated by the two numbers $1$ and $\tau$:
\eqn\torus{
{\Bbb T}= \IC/(\IZ \oplus \tau\IZ)\ , }
in the sense that points in the complex plane are 
identified according to $w \sim w'=w+ n+m \tau$, $n,m\in \IZ$. 
There is, however, a redundancy  in this description of the torus:
the modular parameters $\tau$ and $M \tau$ describe the same torus,
for any modular  transformation $M\in PSL(2,\IZ)$.
The partition function of the theory on this torus is  the trace of the
evolution operator $e^{-H}$ on the finite cylinder of period $l=1$,
described by
  $w=\tau v -u \sim w+n \,,\, 0\le u,v \le 1\,.$
We have 
\eqn\cylh{
H= \tau \, {1\over 2\pi i} \int_0^{-1}\, du\, T(w) -\bar{\tau}
\, {1\over 2\pi i} \int_0^{-1}\, du\, \bar{T}(\bar{w})
}
 and hence,
mapping back to the plane $w\to z=e^{-2\pi i w}$ and using the
transformation law of the energy-momentum tensor \varT, we get
\eqn\torus{Z=\tr_{\CH_P} e^{2\pi i [\tau (L_0-{c\over 24}) -\bar\tau
(\bar L_0-{c\over 24})] } \ }
 with the trace taking care of
the identification of the two ends of the cylinder into a torus.
Using \hilbert\ and the definition \chara\ of characters,
this trace may be written as
\eqn\toruspf{Z=\sum
\N_{j\bar j} \, \chi_j(q) \chi_{\bar j}(\bar q)\qquad
q=e^{2\pi i\tau}\quad \bar q=e^{-2\pi i\bar\tau}
 . }
Let's stress that in these expressions, $\bar\tau$ is the complex conjugate of
$\tau$, and $\bar q$ that of $q$, and therefore, 
$Z=\sum \N_{j\bj} \chi_j(q) \big(\chi_{\bj}(q)\big)^*$ 
is a {\it sesquilinear} form in the characters. 
It is a natural physical requirement that this partition function 
be intrinsically attached to the
torus, and thus be invariant under modular transformations. 
Finally one imposes the extra condition $\N_{11}=1$ which
expresses the unicity of the
identity representation (i.e. of the ``vacuum'').

One is thus led to the problem of
finding all possible sesquilinear forms \toruspf\
with non negative integer coefficients that are modular invariant, and such
that $\N_{11}=1$.
As explained in the previous section, the finite set of characters of any
RCFT, labelled by $\CI$, supports a unitary representation of the
modular group. This implies that any diagonal combination of
characters $Z=\sum_{i\in\CI} \chi_i(q) \chi_i(\bar q)$
is modular invariant. In that case, all representations of 
$\gA\otimes\gA$ appearing in \hilbert\ 
are left-right symmetric, and thus all primary fields are spinless:
$h_j-h_{\bj}=0$. 
This situation is referred to as the ``diagonal case'' or ``diagonal 
theory''. Other solutions are, however, known to exist.

The problem has been completely solved only in a few cases:
for the RCFTs with an affine algebra,
the $\slh(2)$ \CIZK\ and $\slh(3)$ \Ga\ theories at arbitrary level,
plus a host of cases with constraints on the level, e.g. 
the general $\slh(N)$ for $k=1$ \IDG;  
some of the 
associated coset theories \GKO\ have also been fully classified, including
all the minimal $c<1$ theories, $N=2$ ``minimal'' superconformal theories,
etc. A good review on the current
state of the art is provided by T. Gannon \Gannew. For a short account 
of the cases of $\slh(2)$ and $\slh(3)$, see \Zbarilo.

{\petit  \baselineskip=12pt
In the case of CFTs with a current algebra, it is in fact better
to look at the same problem of modular invariants after replacing
in \toruspf\ all specialized characters by non-specialized ones, v.i.z.
$\sum \N_{j\bar j} \chi_j(q,{\bf u}) \big(\chi_{\bj}(q,{\bf u})\big)^* $. 
Because these 
non-specialized characters are linearly independent, there is no ambiguity 
in the determination of the multiplicities $\N_{j\bj}$ from $Z$.
This alternative form of the partition function may be seen to 
result from a modification of the energy-momentum tensor
$T(z)\to T(z) -{2\pi i\over L} ({\bf u},{\bf J}(z))  -{k\over 2} 
\({2\pi\over L}\)^2  ({\bf u},{\bf u})$, see \BPPZ. } 
\medskip

The matrix $Z_{j\bar j}$ of \hilbert\ gives us the spectrum 
of primary fields of the theory. We have also to determine
the couplings of these fields. This may be done by expressing them
in terms of the CVO as
\eqn\primary
{\Phi_{(i,\ib)}(z,\bar{z})=\!\!\! \sum_{j, \jb, k, \bar{k}, t,\bar{t}}\,
\!\!\! d_{(i,\ib) (j,\jb)}^{(k,\bar{k});t, \bar{t}} \,
\phi_{i,j;t}^k(z)\, \otimes
\phi_{\ib,\jb;\bar{t}}^{\bar{k}}(\bar{z})\ .
}
Then matrix elements of $\Phi_{(i,\ib)}$ between  highest weight states
are given in terms of those of CVO, which are supposed to be known
\eqn\threept
{\bra k,\bar k|\Phi_{(i,\ib)}(1,1)|j,\jb\ket= \sum_{t,\bar t}
 d_{(i,\ib) (j,\jb)}^{(k,\bar{k});t, \bar{t}} \,
\bra k|\phi_{i,j;t}^k(1)|j\ket \, 
\bra \bar k|\phi_{\ib,\jb;\bar{t}}^{\bar{k}}(1)|\jb\ket \ . } 
For $\slh(2)$ 
related theories, the latter
have been explicitly computed 
in \refs{\DF,\ZF}.
The chiral $3$-point blocks can be normalised so that
in the diagonal theories the $d$ coefficients are equal to 
a product of Kronecker delta symbols
$\delta_{i \bar i} \delta_{j \bar j} \delta_{k \bar k} \delta_{t \bar t}$
with $t=1, \cdots, \CN_{ij}{}^k$ implying that $d$
vanishes if $\CN_{ij}{}^k= 0$
 and $\sum_{t, \bar{t}}\ d_{(i,i) (j,j)}^{(k,k);t, \bar{t}}=
 \CN_{ij}{}^k\,$.  Thus the  expansion coefficients
$d_{(i,\ib) (j,\jb)}^{(k,\bar{k});t, \bar{t}}\ $ 
give in general the {\it relative} OPE coefficients of
 the non-diagonal model with respect to the  diagonal
 model of  same central charge.
   These numbers are constrained by the requirement of locality
of the physical correlators, which makes use
of the braiding matrices $B(\pm)$.
The resulting set of coupled quadratic equations
has been fully solved only  in the $sl(2)$ cases
(see \refs{\DF,\PZun} and further references therein).

A curious empirical fact was then observed: 
one may introduce graphs, whose properties are intimately 
connected with some features of the  modular invariants or of the 
structure constants. In the simplest case of $\slh(2)_k$ theories, these
graphs are the well-known ADE Dynkin diagrams of Coxeter number $h=k+2$, 
and the set of labels $j$ of the  diagonal terms appearing in $Z$ in 
\toruspf\  is precisely the set of so-called exponents labelling
the eigenvalues $2 \cos \pi j/h$
of the adjacency matrix of the diagram  (see Table 1).  Moreover 
let's introduce the corresponding eigenvectors $\psi_a^{j}$ of the 
adjacency matrix, ($a$ being  a vertex of the diagram $G$) and define the 
structure constants  of the so-called Pasquier algebra \VPdeu
\eqn\pasquiera{ M_{ij}{}^k = \sum_{a\in G} {\psi_a^i \psi_a^j(\psi_a^k)^*\over
\psi_a^1} \ .}
Then the structure constants  $d_{(i,i),(j,j)}{}^{\!\!\!(k,k)}$
of spinless fields in the theory may be shown to be equal  
to those of the Pasquier algebra  $d_{(i,i),(j,j)}{}^{\!\!\!(k,k)}
=M_{ij}{}^k$ \PZun. 
Thus the ADE graph encodes some non trivial information about a subsector
of the theory, namely that of spinless fields, their spectrum and
OPE. 

\medskip

These empirical facts are known to extend to more complicated theories.
In  general, we expect that among the pairs $(j,\bj)$ appearing
in $Z$, a special role will  be played by the diagonal subset
\eqn\expo{\CE=\{(j,\za)| \za=1,\cdots \N_{jj}\} \ ,}
the elements of which, the ``exponents'' of the theory, are thus
counted with the multiplicity $\N_{jj}$.
(In what follows, we shall most of the time omit the multiplicity
index $\za$).
We  assume that $\CE$ is stable under conjugation: $j$ and $j^*$ occur
with the same multiplicity. Then the relevant graphs $G_i$ are labelled
by the set $\CI$, they have a common set of  vertices and their
spectrum is described by the set $\CE$ in the sense that their 
eigenvalues are of the form $S_{ij}/S_{1j}$ and of multiplicity $Z_{jj}$.

The origin and interpretation of these empirical facts had 
remained elusive until recently.  It is one of the 
virtues of BCFT to have cast a new light on these facts and to have 
offered a new framework in which they appear more natural and systematic.

\bigskip

%
\vbox{
\hrule\medskip
\centerline{Table 1: ADE graphs, their Coxeter number and their exponents}
\medskip
\halign{ # & #  &  # & # \cr
 \quad & \qquad\qquad \qquad\qquad \qquad
&\qquad $h$  & \qquad{\rm exponents}\quad  \cr
\cr
  \qquad $A_n$& {\epsfxsize=3cm
        \epsfbox{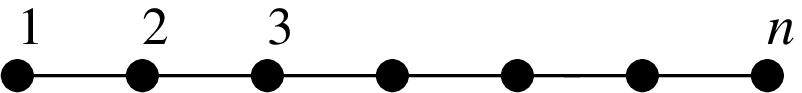}}  & \qquad   $n+1$  &\qquad $1,2,\cdots,n$ \cr
\cr
 \qquad $D_{n+2}$ &\raise -15pt\hbox{\epsfxsize=3cm
        \epsfbox{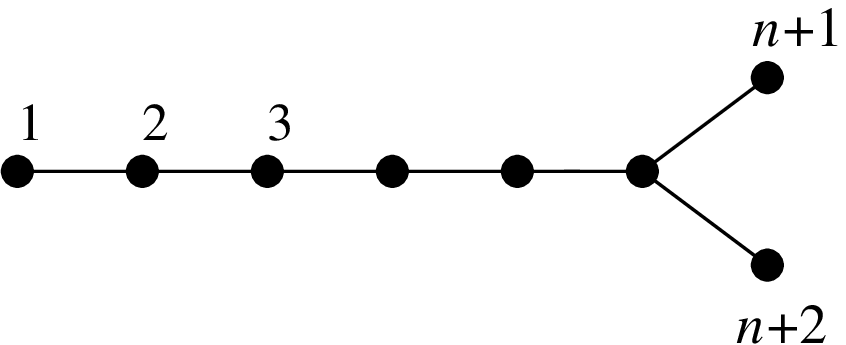}} &\qquad  $2(n+1)$
        &\qquad $1,3,\cdots,2n+1,n+1$ \cr
 \qquad $E_6$ & \raise -5pt\hbox{\epsfxsize=25mm
        \epsfbox{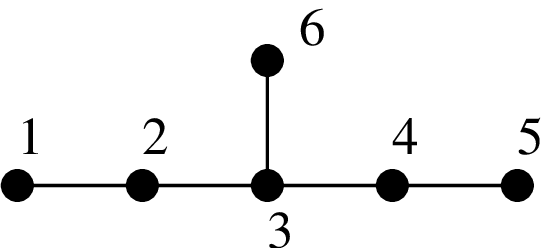}}  &\qquad 12  &\qquad  $1,4,5,7,8,11 $\cr
\cr
 \qquad $E_7$ &  \raise -5pt\hbox{\epsfxsize=3cm
        \epsfbox{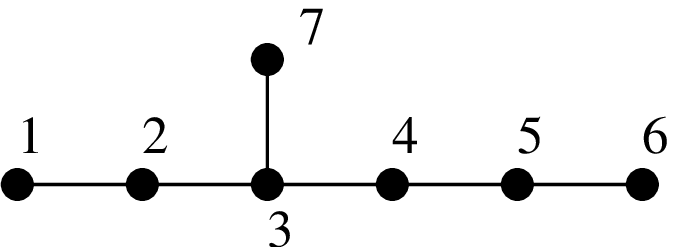}}&\qquad 18& \qquad $1,5,7,9,11,13,17$  \cr
\cr
 \qquad $E_8$ &
\raise -6pt\hbox{\epsfxsize=35mm
        \epsfbox{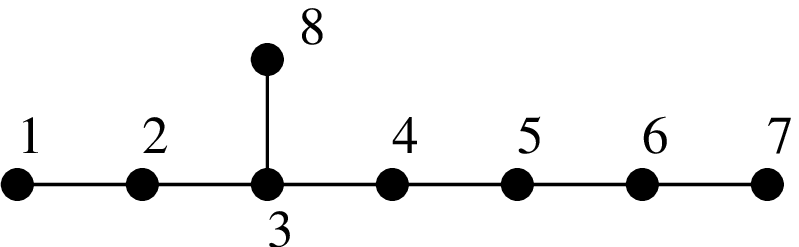}}&\qquad  30   & \qquad $1,7,11,13,17,19,23,29$ \cr
\cr}
\hrule}
\bigskip

%


\newsec{Boundary Conformal Field Theory}
\subsec{Spectral Data in the Upper Half Plane}\nind
We now turn to the study of RCFT in a half-plane. There are
several physical reasons to look at this problem, 
as mentioned in the Introduction.
Here we shall only look at the new information and perspective that
this situation yields on the general structure   of RCFT.

In a half-plane, the admissible diffeomorphisms must respect the
boundary, taken as the real axis: thus only real analytic changes
of coordinates, satisfying $\epsilon(z)=\bar\epsilon(\bz)$ for $z=\bz$ real,
are allowed. The energy momentum itself has this property:
\eqn\bcT{T(z)=\bar T(\bar z)|_{\rm{real\ axis}}\ , }
which expresses simply the
absence of momentum flow across the boundary
and which enables one to extend the  definition of $T$
to the lower half-plane by $T(z):=\bar T(z)$ for $\Im m\, z<0$.
There is thus only {\bf one copy}
 of the Virasoro algebra $L_n=\bar L_n$.
This continuity equation \bcT\ on  $T$  extends
to more general chiral algebras and their generators, at the price
however of some complication. In general, the continuity
equation on generators of the chiral algebra involves some
automorphism of that algebra:
\eqn\bcW{ W(z)=\Omega\bar W(\bar z)|_{\rm{real\ axis}}}
(see \BPPZ\ and further references therein).

The half-plane, punctured at the origin, (which introduces a distinction
between the two halves of the real axis), may also be conformally mapped
on an infinite horizontal strip of width $L$ by $w={L\over \pi} \log z$.
Boundary conditions, loosely specified at this stage by labels $a$ and
$b$, are assigned to fields on the two boundaries $z$~real~$> 0$, $< 0$
or $\Im m\, w=0, L$.
For given boundary conditions on the generators of the algebra
and on the other fields of the theory, i.e. for given
automorphisms $\Omega$ and given $a,b$,
we may again use a description of the system
by a Hilbert space of states $\CH_{ba}$ (we drop the dependence on $\Omega$
for simplicity). 
This space decomposes on representations of Vir or $\gA$ according to
\eqn\Hab{ \CH_{ba}= \oplus n_{ib}{}^a \CV_i\ ,}
with  a new set of multiplicities $ n_{ib}{}^a\in{\IN}$.
The natural Hamiltonian on the strip is 
the translation operator in $\Re e\, w$, hence, mapped back in the half-plane
\eqn\Hstrip{ H_{b|a}={\pi\over L}\left(L_0-{c\over 24}\right)\ . }

To recapitulate, in order to fully specify the operator content of the theory
in various configurations, we need not only determine the multiplicities
``in the bulk''$ \N_{j\bar j}$ of \hilbert, but also the possible
boundary conditions $a,b$ on a half-plane and the associated multiplicities
$n_{ib}{}^a $. This will be our task in the following, and as we shall
see, a surprising fact is that the latter have some bearing on the former.

\fig{ The same domain seen in different coordinates: a semi-circular 
annulus, with the two half-circles identified, a rectangular 
domain with two opposite sides identified, and a circular annulus. }
{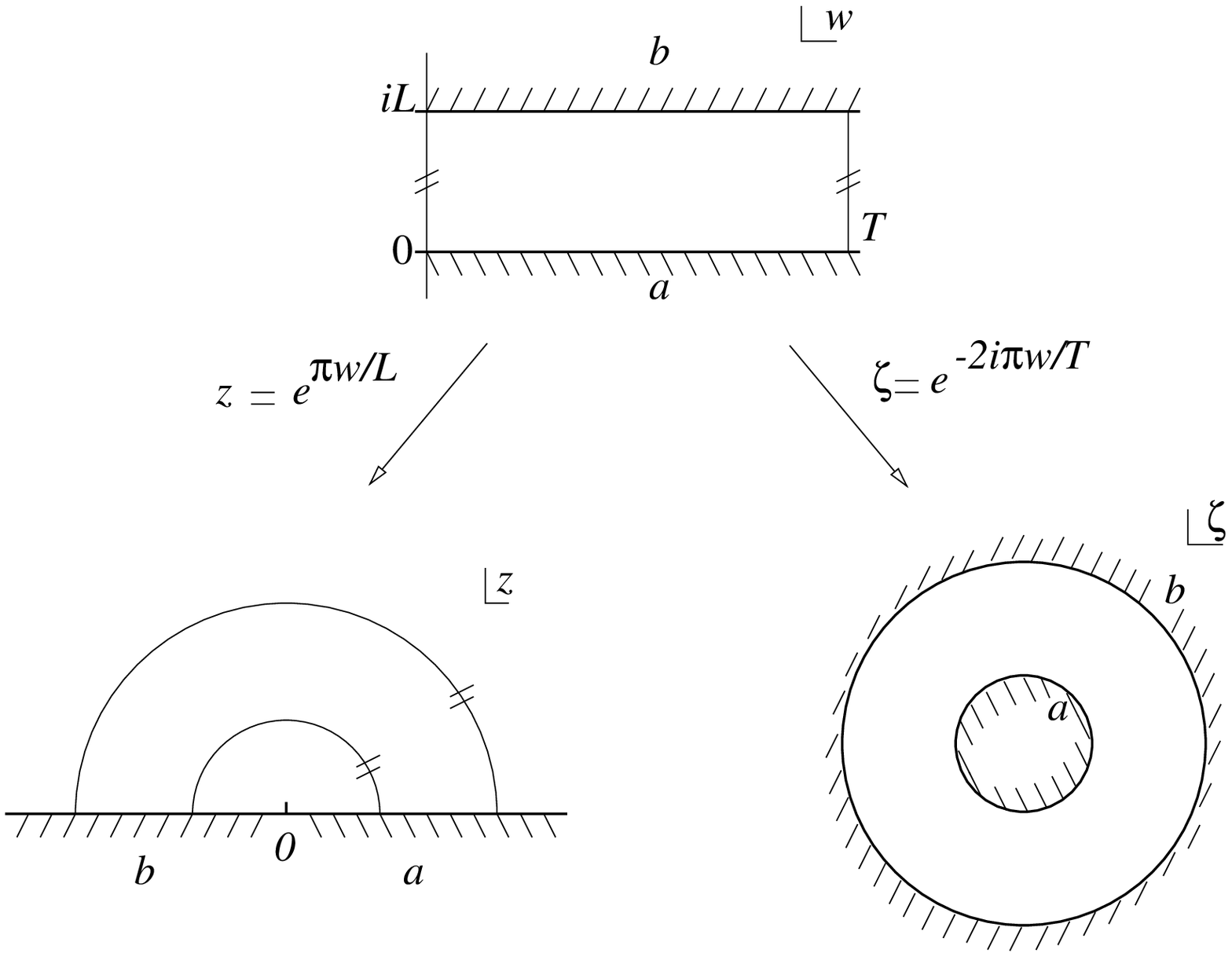}{4.0truein}\figlabel\annulus

\subsec{Boundary states}
\nind In the same way that we found useful to chop a finite segment of the
infinite cylinder and identify its ends to make a torus, it is
suggested to consider a finite segment of the strip -- or a
semi-annular domain in the half-plane-- and identify its edges,
thus making a cylinder. 
This cylinder can be mapped back into an annular domain in the plane,
with open boundaries.
More explicitly, consider the segment $0\le \Re e\,w\le T$ of the strip
--i.e. the semi-annular domain in the upper half-plane
comprised between the semi-circles of radii 1 and $e^{\pi T/L}$, the latter
being identified. It may be
conformally mapped into an annulus in the complex plane
by $\zeta=\exp(- 2i\pi w/T)$, of radii $1$ and $e^{2\pi L/T}$, see Fig. 
\annulus. By
working out the effect of this change of coordinates on the energy-momentum
$T$, using \varT, one finds that \bcT\ implies
\eqn\newbcT{ \zeta^2 T(\zeta)= \overline{\zeta}^2 \overline{T}(\bar \zeta)
\quad{\rm for}\ |\zeta|=1,\ e^{2\pi{L\over T}}\ .}
After radial quantization, this translates into
 a condition on {\it boundary states} $|a\rangle\, ,\ |b\rangle\, \in \CH_P$
which describe the system on these two boundaries. 
\eqn\bcsta{(L_n-\bar L_{-n})|a\rangle=0 }
(and likewise for $|b\ket$). 
{\petit  \baselineskip=12pt
\Ex : assuming that $W$ transforms as a primary field of
conformal weight $(h_W,0)$, find the corresponding condition on
$W(\zeta)$. 
Then show that the analogue of \bcsta\ reads
$(W_n- (-1)^{h_W} \Omega(\bar W_{-n}))|a\rangle=0$.}

\medskip
We shall now look for a basis of states, solutions of this linear
system of boundary conditions.
One may seek solutions of these equations in each
$\CV_j\otimes \CV_{\bar j} \subset \CH_P$,
since these spaces are invariant under the action of the two copies
of Vir or of the chiral algebra $\gA$.  Consider only for simplicity
the case of the Virasoro generators.

\smallskip
\nind {\bf Lemma}\ {\sl
There is an independent ``Ishibashi state'' $|j\rrangle$, 
solution of \bcsta,  
for each $j=\bar j$, i.e. $j\in \CE$, the set of exponents.}
\smallskip\nind
Proof (G. Watts)\Wa: Use the identification between states
$ |a\ket \in \CV_j\otimes \CV_{\bar j}$  
and operators $X_a\in \hbox{Hom}(\CV_{\bar j}, \CV_{j})$,
namely
 $|a\ket=\sum_{n,\bar n} a_{n,\bar n} |j,n\ket \otimes |\bar j, \bar n\ket $
$\leftrightarrow 
X_a =\sum_{n,\bar n} a_{n,\bar n} |j,n\ket \bra\bar j, \bar n| $.
Here we make use of the scalar product in $\CV_{\jb}$ for
which $\bar L_{-n}  =\bar L_n^\dagger $, hence 
\bcsta\ means that $L_n X_a=X_a L_n$, i.e.
$X_a$ intertwines the action of Vir in the two \irrep s
$\CV_j$ and $ \CV_{\bar j}$. By Schur's lemma, 
this implies that they are 
equivalent,  $\CV_j \sim \CV_{\bar j}$, i.e. that their labels
coincide $j=\bar j$ and that $X_a$ is proportional to $P_j$, the projector
in $\CV_j$. We shall denote $|j\rrangle$ the corresponding state,
solution to \bcsta. \par\nind
{\petit  \baselineskip=12pt
Since ``exponents'' $j\in \CE$ may have some
multiplicity, an extra label should be appended to our
notation $|j\rrangle$. We omit it for the sake of simplicity.
The previous considerations extend with only notational complications
to more general chiral algebras and their possible
gluing automorphisms $\Omega$. See \BPPZ\ for more details and more
references on these points.  Also, see \FS\ for an alternative  
discussion of Ishibashi states.  }
\bigskip

The normalization of this ``Ishibashi state'' requires some care.
One first notices that, for $\tilde q$ a real number between 0 and 1,
\eqn\chIsh{\llangle j'| \tilde q^{\oh(L_0+\bar L_0 -{c\over 12})}|j\rrangle
=\delta_{jj'}{\chi_j(\tilde q)}}
up to a constant that we choose equal to 1.
It would seem natural to then define the norm of these states by
the limit $\tilde q\to 1$ of \chIsh. This limit diverges, however, and
the adequate definition is rather
\eqn\normIsh{\llangle j|\!|j'\rrangle =
\delta_{jj'} {S_{1j}}}
{\petit  \baselineskip=12pt
This comes about in the following way: a natural
regularization of the above limit is:
\eqn\newnorm{\llangle j|\!|j'\rrangle =
\hbox{lim}_{\tilde q\to 1}
q^{c/24} \llangle j'| \tilde q^{\oh(L_0+\bar L_0 -{c\over 12})}
|j\rrangle}
where $q$ is the modular transform of $\tilde q=e^{-2\pi i/\tau}$,
$q=e^{2\pi i\tau}$. In a (``unitary'') theory in which the identity
representation (denoted 1) is the one with the smallest conformal weight,
show that in the limit $q\to 0$, the r.h.s. of  \newnorm\ reduces
to \normIsh. In non unitary theories, this limiting procedure fails, 
but we keep \normIsh\ as a definition of the new norm.}
\medskip

At the term of this study, we have found a basis of solutions to
the constraint \bcsta\ on boundary states, and it is thus legitimate
to expand the two states attached to the two boundaries of our domain
as
\eqn\expbs{ |a\rangle =\sum_{j\in \CE} {\psi_a^j \over \sqrt{S_{1j}}}
\;|j\rrangle }
with coefficients denoted $\psi_a^j$, and likewise for 
$|b\rangle$. We define an involution $a\to a^*$ 
on the boundary states by $\psi_{a^*}^j=\psi_a^{j^*}=(\psi_a^{j})^*$, 
(recall that $j\to j^*$ is an involution in $\CE$). One may show 
\RS\ that it is natural to write for the conjugate state
\eqn\expconjs{\langle b| =\sum_{j\in \CE}  \llangle  j|\,
{\psi_{b^*}^j \over \sqrt{S_{1j}}}\ .}
 As a  consequence
\eqn\scalpr{
\bra b\|a\ket
=\sum_{j\in \CE}{\psi_a^j\left(\psi_b^j\right)^*\over S_{1j}}
\llangle j \| j\rrangle =
\sum_{j\in \CE}
{\psi_a^j \left(\psi_b^j\right)^*} }
so that  the orthonormality of the boundary states is equivalent  to
that of the $\psi$'s.

\subsec{Cardy equation}
\nind Let us return 
to the annulus $ 1\le |\zeta|\le e^{2\pi{L/ T}}$
considered in last subsection, or equivalently 
to the cylinder of length $L$ and perimeter $T$, 
with boundary conditions (b.c.)
$a$ and $b$ on its two ends. Following Cardy \Cabc, we shall
compute its partition function $Z_{b|a}$ in two different ways.
If we regard it as resulting from the evolution between the
boundary states $|a\ket$ and $\bra b|$, with $\tilde q^\oh=e^{-2\pi{L/T}}$, 
we find
\eqn\Zabun{\eqalign{
Z_{\b|\a}&= \langle
\b| ({\tilde q }^{\oh(L_0+\bar L_0 -{c\over 12})}
|\a\rangle              
=\sum_{j,j'\in\CE}
{\left(\psi_b^j\right)\!{}^*\,\psi_a^{j'} \over S_{1j}}
\llangle j|\tilde{q}^{\,{1\over 2}(L_0+\overline{L}_0-{c\over 12})}
|j'\rrangle\cr
&= \sum_{j\in\CE}
\psi_a^j \left(\psi_b^j\right)\!{}^*\, {\chi_j(\tilde{q})\over S_{1j}}\ .\cr
} }

\fig{Two  alternative computations of the partition
function $Z_{b|a}$: (a) on the cylinder, between the
boundary states $|a\rangle$ and $\langle b|$,  (b) as a periodic time
evolution on the strip, with boundary conditions $a$ and $b$.}{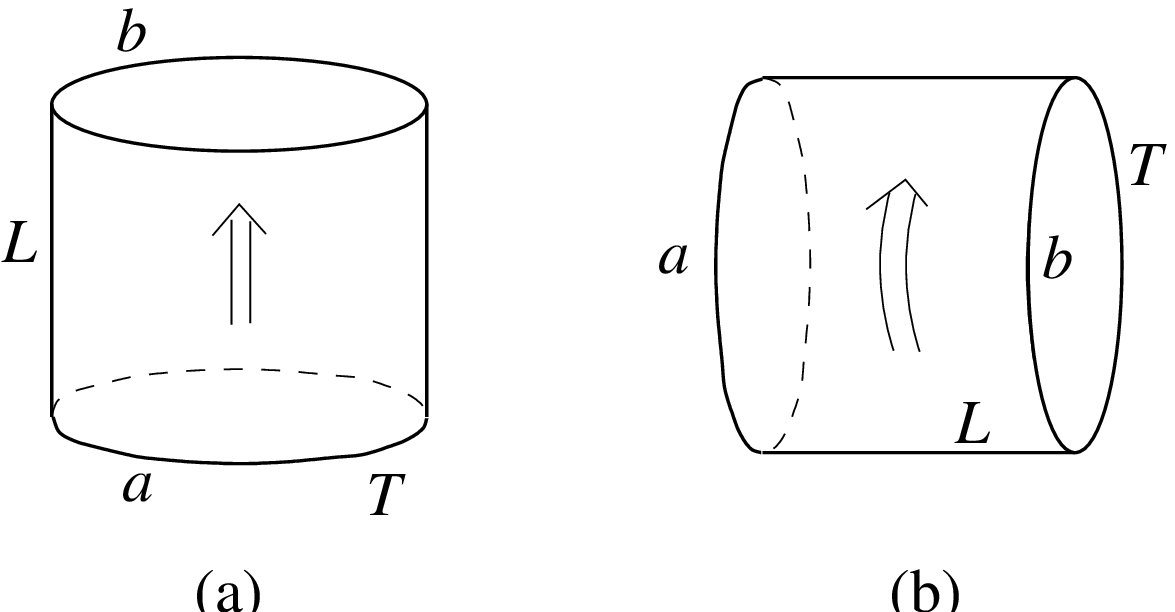}
{8cm}\figlabel\cardyeq

On the other hand, if we regard it as resulting from the
periodic  ``time''
evolution on the strip with b.c. $a$ and $b$, using
the decomposition \Hab\ of the Hilbert space $\CH_{ba}$,
and with $q=e^{-\pi T/L}$
\eqn\Zabde{
Z_{b|a}(q)=\sum_{i\in\CI} n_{ib}{}^a \chi_i(q)\ .}
See Fig. 2. Note that string theorists would refer to these two situations as 
(a): the tree approximation of the propagation of a closed string; 
(b) the one-loop evolution of an open string. 
Performing a modular transformation on the characters
$\chi_j(\tilde q)=\sum_i S_{ji^*}\chi_i(q)$ in \Zabun,
and identifying the coefficients of $\chi_i$ yields
\eqn\cardy{
n_{ia}{}^b =\sum_{j\in\CE} \, {S_{ij}\over S_{1j}}\,
\psi_a^j \left(\psi_b^j\right)\!{}^*  \ ,}
a fundamental equation for our discussion that we refer to as Cardy 
equation \Cabc.
In deriving this form of Cardy equation, we have made use
of a symmetry property of $n_{ia}{}^b$
\eqn\symnab{
n_{i\a}{}^\b =n_{i^*\b}{}^{\a} 
\ }
which follows from  the symmetries of $S$. 
\nind{\petit  \baselineskip=12pt
(Comment: this identification of coefficients of specialized
 characters is in general not justified, as the $\chi_i(q)$ are not linearly
independent. As in sect. 1, it is better to 
generalize the previous discussion, in a way which introduces 
non-specialized --and linearly independent-- characters. This has
been done in \BPPZ\ for the case of CFTs with a current algebra.
Unfortunately, little is known about non-specialized characters
for other chiral algebras.)}
\medskip

Let us stress that in \cardy, the summation runs over $j\in \CE$, i.e.
this equation incorporates some information on the spectrum of the 
theory ``in the bulk'', i.e. on the modular invariant partition function 
\toruspf.
 
Cardy equation \cardy\ is a non linear constraint relating
a priori unknown complex coefficients $\psi^j$ to unknown 
integer multiplicities
$n_{ia}{}^b$. We need additional assumptions to exploit it.

We shall thus assume that
\item{$\bullet$} we have found an orthonormal set  of boundary states
$|a\rangle$, i.e. satisfying
\eqn\ortho
{ (n_1)_a{}^b=\sum_{j\in\CE}\psi_a{}^j(\psi_b{}^j)^*=
\delta_{ab}\ ;}
\item{$\bullet$} we have been able to construct a  {\it complete} set
 of such boundary states $|a\rangle$
\eqn\compl{ \sum_a \psi_a{}^j(\psi_a{}^{j'})^*=\delta_{jj'} \ .}
\par\nind
These assumptions imply that
$$ \#\ \hbox{boundary states}=\#\ \hbox{independent Ishibashi states}=|\CE|\ .$$
None of these assumptions is innocent.  In particular, 
it is in general not consistent to assume that all boundary states 
are mutually orthogonal; for example it has been shown recently that  the 
renormalisation group flow may bring a theory to a boundary state 
where orthogonality is violated \refs{\Affleck,\Reck,\Watts}. 
Such a state $A$ is a linear superposition with non-negative integer 
coefficients of boundary states satisfying \ortho, see the exercise
 below.   As for the assumption
of completeness, it is not obviously natural  and
would also deserve a critical discussion; relaxing it would 
force us to reconsider the forthcoming discussion.

{\petit Exercise:
Suppose we have found a complete and orthonormal solution $\{\psi\}$ to
\cardy\ and consider a boundary state $|A\ket=\sum_{j\in\Exp} 
\Psi_A^j (S_{1j})^{-1/2}\, 
|j\rrangle$ which also satisfies \cardy\ with itself and with all
the orthonormal states $|a\ket$: $\sum_{j\in \Exp} \Psi_A^j
\psi_a^{j*} S_{ij}/{S_{1j}} =N_{iA}{}^a\in \IN$. Prove that this 
implies that 
\eqn\intPs{\Psi_A{}^j=\sum_a N_{1A}{}^a \psi_a^j\ ,}
and that conversely, the integrality of $N_{iA}{}^a$ and of 
$N_{iA}{}^A$ follows from \intPs, with $N_{1A}{}^a\in \IN$.
}


\subsec{Representations of the fusion algebra and graphs}
\nind
Return to Cardy equation \cardy\ supplemented by the above assumptions
\ortho-\compl\   and observe that it gives a
decomposition of the matrices $n_i$, defined by $(n_i)_a{}^b=n_{ia}{}^b$,
into their orthonormal eigenvectors $\psi$ and their eigenvalues
$S_{ij}/S_{1j}$. Observe also that as a consequence of Verlinde formula 
\verl, these eigenvalues form a one-dimensional representation of
the fusion algebra
\eqn\onedrep{
{S_{i\ell}\over S_{1\ell}}\; {S_{j\ell}\over S_{1\ell}}
=\sum_{k\in\CI} \CN_{ij}{}^k\; {S_{k\ell}\over S_{1\ell}}\, , 
\qquad \forall i,j,\ell\in \CI\ .} 
Hence the matrices $n_i$ also form a representation
of the fusion algebra \regfa
\eqn\nfus
{n_i\;n_j=\sum_{k\in\CI} \CN_{ij}{}^k\,n_k \ }
and they thus commute. Moreover, as we have seen above, 
they satisfy $n_1=I$, $n_i^T=n_{i^*}$.

Conversely, consider any ${\IN}$-valued matrix representation of 
the Verlinde fusion
algebra $n_i$, such that $n_i^T=n_{i^*}$. Since the algebra is
commutative, $[n_i,n_i^T]=[n_i,n_{i^*}]=0$, the $n_i$ 
commute with their transpose, ({\it normal matrices}), hence 
they are diagonalizable in a common orthonormal basis. Their
eigenvalues are known to be of the form $S_{ij}/S_{1j}$. They 
may thus be written as in \cardy. Thus any such ${\IN}$-valued 
matrix representation of the Verlinde fusion algebra
 gives a (complete orthonormal) solution to Cardy's equation.

\medskip\nind 
{\bf Conclusion:}\
\medskip

\encadre{
$${\IN}\hbox{-valued matrix representation  of the fusion algebra, } 
n_i^T=n_{i^*}$$
$$\Longleftrightarrow \hbox{  Complete, orthonormal solution of Cardy equation}$$
}

\smallskip
\nind
Moreover, since ${\IN}$-valued matrices are naturally interpreted as graph
adjacency matrices,  graphs appear naturally and their spectral properties 
are those anticipated in the last lines of sect. 1. 

{\petit  \baselineskip=12pt
The relevance of the fusion algebra in the solution of Cardy equation
had been pointed out by Cardy himself for diagonal theories \Cabc\ 
and foreseen  
in general in \DFZun\ with no good justification; the importance of the
assumption of completeness of boundary conditions
 was first stressed by Pradisi et al \PSS.}


\subsec{The case of $\widehat{sl}(2)$ WZW theories}
\noindent
{\bf Problem:}\
{\sl Classify all ${\IN}$-valued matrix reps of $\widehat{s\ell}(2)_k$ fusion
algebra with $k$ fixed.}

\noindent The algebra is generated recursively by $n_2$
\eqn\recurr{n_1=I, \qquad n_2\;n_i=n_{i+1}+n_{i-1},\quad i=2,\ldots,k}
$$
\hbox{$S$ real}\ \Rightarrow\  n_i=n_i^T \ .
$$

\noindent
Even though $\psi^j$ and $\CE$ are a priori unknown,  
we know from \Ssld\ that $n_2$ has eigenvalues of the form
\eqn\evalu{
\gamma_j={S_{2j}\over S_{1j}}=2\cos{\pi j\over k+2},\quad j\in\CE \ .
}
We shall discard the case where the matrix $n_2$ is ``reducible'', 
i.e. may be written as a direct sum of two matrices 
$n_2=n_2^{(1)}\oplus n_2^{(2)}$, because then all the other matrices 
$n_i$ have the same property and this  corresponds to decoupled 
boundary conditions. It turns out that  
irreducible  ${\IN}$-valued matrices $G$
with spectrum $|\gamma|<2$ have been classified \GHJ. They are
the adjacency matrices  either of the $\hbox{$A$-$D$-$E$}$ Dynkin
diagrams or of the ``tadpoles'' $T_n={A_{2n}/{\IZ}_2}$. 
Thus as a consequence of equation \cardy\ alone, 
 for a $\slh(2)$ theory at level $k$, the possible boundary conditions
are in one-to-one correspondence with the vertices of one of 
these diagrams $G$, with Coxeter number $h=k+2$. If we remember, however,
 that the set $\CE$  must appear in one of the modular invariant torus
partition functions, the case $G=T_n$ has to be discarded, and we are 
left with $ADE$. (Up to this last step, this looks like the 
simplest route leading to the $ADE$ classification of $\slh(2)$ theories.)
We thus conclude that for each $\slh(2)$ theory classified by a 
Dynkin diagram $G$ of $ADE$ type
\eqnn\conclu
$$\eqalign{ \CE &=\hbox{Exp}(G),\qquad \hbox{dim}(n_i)=|\CE|=|G| \cr
\hbox{complete } & \hbox{ orthonormal b. c. } = a,b,\cdots\ :\ 
\hbox{vertices of $G$} \cr
n_2&=\hbox{ adjacency matrix of $G$ }\cr
n_i&=\hbox{``$i$-th fused adjacency matrix'' of $G$ }\cr
\psi^j &=\hbox{eigenvector of $n_2$ with eigenvalue $\gamma_j$}\ . 
}
$$
One checks indeed that 
the matrices $n_i$, given by equation \cardy, together with 
\Ssld, have only non negative integer elements. 
See \refs{\BPPZ,\Zbarilo} for a review of their remarkable 
properties and of their
ubiquitous r\^ole in a variety of problems.


\subsec{The case of $c<1$ minimal models}
\nind
As recalled above, this case is closely related to the $\slh(2)$ models
that we just discussed.    
If $c=1-{6(p-p')^2\over pp'} $, as in sect 1.1, 
the classification of modular invariants is done by
pairs of Dynkin diagrams $(A_{p'-1},G)$, with $p$
equal to the Coxeter number of $G$.
Our problem is then to classify all ${\IN}$-valued matrix representations 
of the corresponding fusion algebra.
\smallskip
\noindent
{\bf Theorem:} \BPPZ\ {\sl The only complete orthonormal solution to 
Cardy's equation are 
labelled by pairs $(r,a)$ of nodes of the $A_{p'-1}$ and of
the $G$ graphs, with the identification
\eqn\kacsymm{
(r,a) \equiv (p'-r,\gamma(a)) 
}
where $\gamma$ is the following automorphism of the $G$ Dynkin diagram:
the natural  ${\IZ}_2$ symmetry for the $A$, $D_{{\rm odd}}$
and $E_6$ cases, the identity for the others.}
\def\nn{n^{(\slh(2))}}
\def\nn{n^{(G)}}
\par
\nind
The solutions are given explicitly as
\eqn\bcmin
{n_{rs}=N_r\otimes \nn_s + N_{p'-r}\otimes \nn_{p-s} }
or
\eqn\bcMin{
n_{rs; (r_1,a)}{}^{(r_2,b)}=N_{rr_1}{}^{r_2}\, \nn_{sa}{}^b
+N_{p'-r\,r_1}{}^{r_2}\, \nn_{p-s\,a}{}^b\ , 
}
with $1\le r,r_1,r_2 \le p'-1$, $1\le s\le p-1$, and
$a,b$ are running over the nodes of the Dynkin diagram $G$;
$N_r$ are the fusion matrices of $\widehat{sl}(2)_{p'-2}$,
$\nn_s$ are the representation matrices of the fusion algebra of
$\widehat{sl}(2)_{p-2}$ associated with the graph $G$
and introduced in sect 2.5.
\smallskip
\noindent {\it Hint:} Look at the generators $n_{21}$
and $n_{12}$ of fusion algebra. Both have eigenvalues
$|\gamma|<2$, but now they are no longer irreducible matrices
as in sect. 2.5. To show that \bcmin\ is indeed the only 
solution requires a detailed discussion, see \BPPZ. 

\smallskip
\noindent{\petit \baselineskip=12pt
 Example: The 3-state Potts model. 
This is the $p'=5$, $p=6$
($c=4/5$) minimal model, classified by the pair $(A_4,D_4)$.
According to the previous discussion, there are 8 distinct 
boundary conditions labelled by
$$(r,a)\in (A_4,D_4)/{\IZ}_2  
\qquad {\rm e.g.}\ r=1,2,\ a\in D_4
$$
 This was first pointed out in \AOS: the three boundary conditions 
$(1,a)$ where $a=1,3,4$ is one of the three end-points of the $D_4$ diagram
are fixed b.c., corresponding  to fixing the value of the Potts 
``spin'' to one of its three values, while in the three b.c. $(2,a)$, 
the boundary spin may take either value different from $a$; the 
b.c. $(1,2)$, where $2$ denotes the middle point of the diagram,
is the free boundary condition: the boundary spin may take an 
arbitrary value; finally the b.c. $(2,2)$ is more delicate to
describe \AOS. The explicit expressions of the  different partition 
functions of type $Z_{(1,1)|(r,a)}$ read
\eqnn\Zpotts
$$\eqalignno{Z_{(1,1)|(1,1)}&=\chi_{(1,1)}+\chi_{(1,5)} \cr
Z_{(1,1)|(1,2)}&=\chi_{(4,2)}+\chi_{(4,4)}    \cr
Z_{(1,1)|(1,a)}&=\chi_{(1,3)}   \quad {\rm if\ } a=3,4 \cr
Z_{(1,1)|(2,1)}&=\chi_{(3,1)}+\chi_{(3,5)} \cr
Z_{(1,1)|(2,2)}&=\chi_{(2,2)}+\chi_{(2,4)}    \cr
Z_{(1,1)|(2,a)}&=\chi_{(3,3)}   \quad {\rm if\ } a=3,4\ . 
}$$
The others are linear combinations with non-negative integer coefficients
of the latter. See also \refs{\BPZ,\BPde}
 for a lattice realization with integrable boundary weights.  } 


\subsec{Other cases}
\nind
It should be clear that the situation that we have described 
in detail for $sl(2)$ extends to all RCFTs. 
The matrices $n_i$, solutions to Cardy equation, 
are the adjacency matrices of graphs. 
In the case of $\slh(N)$, it is sufficient to supply the $(N-1)$ 
fundamental matrices $\npbox$, 
$p=1,\cdots, N-1$,  indexed by the Young tableaux of the
 fundamental representations of $sl(N)$,  to determine
all of them. The fact that  all $n_i$ have non negative integer 
elements is then non trivial. 
By Cardy equation again, they satisfy a very restrictive
spectral property: their eigenvalues must be of the form $S_{ij}/S_{1j}$,
when $j$ runs over the set $\CE$, i.e. the diagonal part of the modular
invariant. 

The program of classifying these graphs/boundary conditions 
has been completed only in a few cases: $\slh(2)$ as discussed
above, $\slh(3)$ through a
combination of some empirical search of relevant graphs \refs{\DFZdeu\BPPZ},
 of Gannon's classification of the modular invariants \Ga,  
  and of the recent work of Ocneanu \AO, 
see \refs{\BPPZ,\Zbarilo} for a discussion;
$\slh(N)_1$ \BPPZ,   where the results match those obtained in 
the study of modular invariants \IDG: the graphs turn out to be star 
polygons.  


\newsec{Boundary Operator Algebra}
\nind According to Cardy \Cabc, changes of boundary conditions 
can be interpreted as due 
to the insertion of fields ${}^b\Psi^a_{j,\zb}(x)\,$  living on
the boundary, $\Im m\, z=0$, $x=\Re e\, z$ of the upper half-plane 
$z\in  {\Bbb H}$.
We know the spectrum of these fields from the 
previous discussion: for a given pair $a,b$ of boundary conditions and 
a label $j\in \CI$, there are $n_{ja}{}^b$ independent such fields, 
which are thus labelled by a multiplicity label $\za=1, \cdots, n_{ja}{}^b$.
Pictorially, we may again use a CVO-like representation for
 ${}^b\Psi^a_{j,\za}(x)\,=$\hbox{\raise -2mm\hbox{
 \epsfxsize=10mm\epsfbox{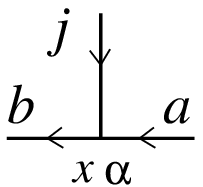}}}. 
It is a natural --and physically important-- question to determine
the correlation functions of these new fields in the possible presence 
of the ``usual'' fields ``in the bulk''. 
Two quantities of particular importance are 
\item{i)} the fusion matrix  $\Fo$ of boundary operators, 
which  plays for the boundary fields the same
r\^ole as the matrix  $F$ for the CVO (sect.~1.1), 
\hbox{\raise -2mm\hbox{\epsfxsize=45mm\epsfbox{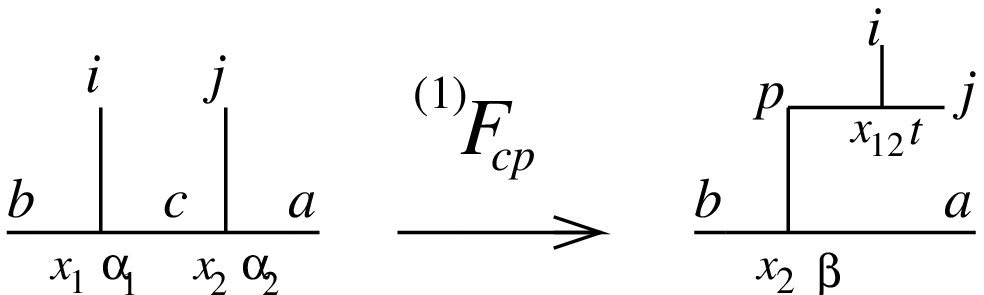}}},
gives their OPE coefficient  
\eqn\bOPE
{\!\!\!\! {}^{b}\Psi_{i, \za_1}^c(x_1)\, {}^{c}\Psi_{j, \za_2}^a(x_2)\,
=
\sum_{p,\zb, t }\
{}^{(1)}F_{c p}\left[\matrix{
i&j\cr b&a} \right]_{\za_1\, \za_2}^{\zb\  \ t}\
\!\!\!\!\!\!\bra p,0|\phi^p_{ij;t}(x_{12})|j,0\ket \
{}^{b}\Psi_{p, \zb}^a(x_2) + \  \dots \,}
in which the multiplicity labels run over $\za_1=1,\cdots, n_{ic}{}^b$, 
$\za_2=1,\cdots, n_{ja}{}^c$, $\zb=1,\cdots, n_{pa}{}^b$, $t=1,\cdots, 
\CN_{ij}{}^ p$; note that eq. \nfus\ is again a necessary condition 
for the matrix $\Fo$ to be invertible;
\item{ii)} the bulk-boundary ``reflection'' coefficients
$ R_{a,\za}^{(i,\bar i^*,t)}(p) $ 
(denoted ${}^{a, \za}B_{(i, \bar{i})}^{p, t}$ in \BPPZ)
enable one to expand bulk fields $\Phi$ in terms of $\Psi$, 
close to the boundary, i.e. for small $y=\Im m\, z$ 
\eqn\bulkb{\Phi_{(i, \bar{i})}(z\,,\bar{z})
= \sum_{a,\za\,, p\in \CI\,, t}\ 
 R_{a,\za}^{(i,\bar i^*,t)}(p)
\bra p,0|\phi^p_{i\bar{i};t}(2iy)|j,0\ket \,
{}^{a}\Psi_{p, \za}^a(x) + ...
}
Here $\za=1,\cdots,n_{pa}{}^a$, $t=1,\cdots,\CN_{i\bar i}{}^{p^*}$. 
Given these data (and for a chosen normalisation of the $\Psi$), 
we may in principle compute all correlation functions of 
$\Psi$ and $\Phi$ \CaL.

 We shall not dwell here on the determination of the fusion matrices 
$\Fo$ and of the bulk-boundary
coefficients. They have been the object of much activity lately, 
in particular on their connection with chiral data and 
on the algebraic relations that they satisfy.
Their explicit calculation for the $A$ or $D$ type minimal
models has been completed by Runkel \refs{\Ru,\Rudeu}. 
Let us just mention two results:\par
In the diagonal theories, the matrices $\Fo$ coincide with the 
standard fusion matrices $F$: this 
observation first made in \Ru\ for the $sl(2)$ case through a
tedious computation is in accordance with the fact
that in that case, the indices $a,b,\cdots$ are of the same nature
as $i,j,\cdots$, i.e. also belong to the set $\CI$;
this result was generally established in \BPPZ\ rewriting the boundary field
Lewellen  sewing equation as a pentagon relation, see Fig. 3 below.
Also, the reflection  coefficients $R(p)$ are expressible in terms
of the matrix $S(p)$ introduced at the end of sect. 1.1, a quite remarkable 
convergence between seemingly very different objects.   
\par
In general, the bulk-boundary coefficient pertaining to the identity, i.e.
$ R_{a,1}^{(i, i^*,1)}(1)$ is proportional to $\psi_a{}^i/\psi_a{}^1$. 
It thus satisfies, up to a normalisation, the Pasquier algebra \pasquiera,
$$ {\psi_a{}^i\over\psi_a{}^1}{\psi_a{}^j\over\psi_a{}^1}=
\sum_{k\in \CI}\,M_{ij}{}^k\, {\psi_a{}^k\over\psi_a{}^1} \,.  $$
This expresses the OPA  of bulk fields of type $\Phi_{(i,i)}$ 
near the boundary $(a)$ \refs{\PSS,\RS,\FS,\BPPZ}, 
and  must be compared with the 
empirical observation \PZun\  mentioned in sect. 1.2 on the r\^ole of 
the Pasquier algebra in the OPA of spinless  fields,
now derived in BCFT.

\medskip

The determination of the $\Fo$ matrix, whose entries are 
called ``cells'' by Ocneanu \Ocn, turns out to be an essential
step, not only in the study of boundary effects, but also in
uncovering hidden algebraic aspects of the theory (sect.~5)
and in the study of associated lattice models (sect.~6). 
For a given set of matrices $\{n_i\}$ --a given ``graph''--
this set of cells satisfies various non linear relations:
orthonormality expressing the fact that it plays 
the r\^ole of a change of basis (``$3j$-symbols'') 
in tensors products in a certain space, 
a mixed pentagon identity written  symbolically  as 
$ F\, \Fo\, \Fo= \Fo\, \Fo$, 
 expressing the consistency (associativity) in the 
fusion of several boundary fields, (see Fig. 3),  and other identities. 
See \refs{\AO,\PZtmr,\PZnew} for more details and references. 
\fig{  The mixed pentagon identity expressing the associativity of the 
fusion of boundary fields.  } {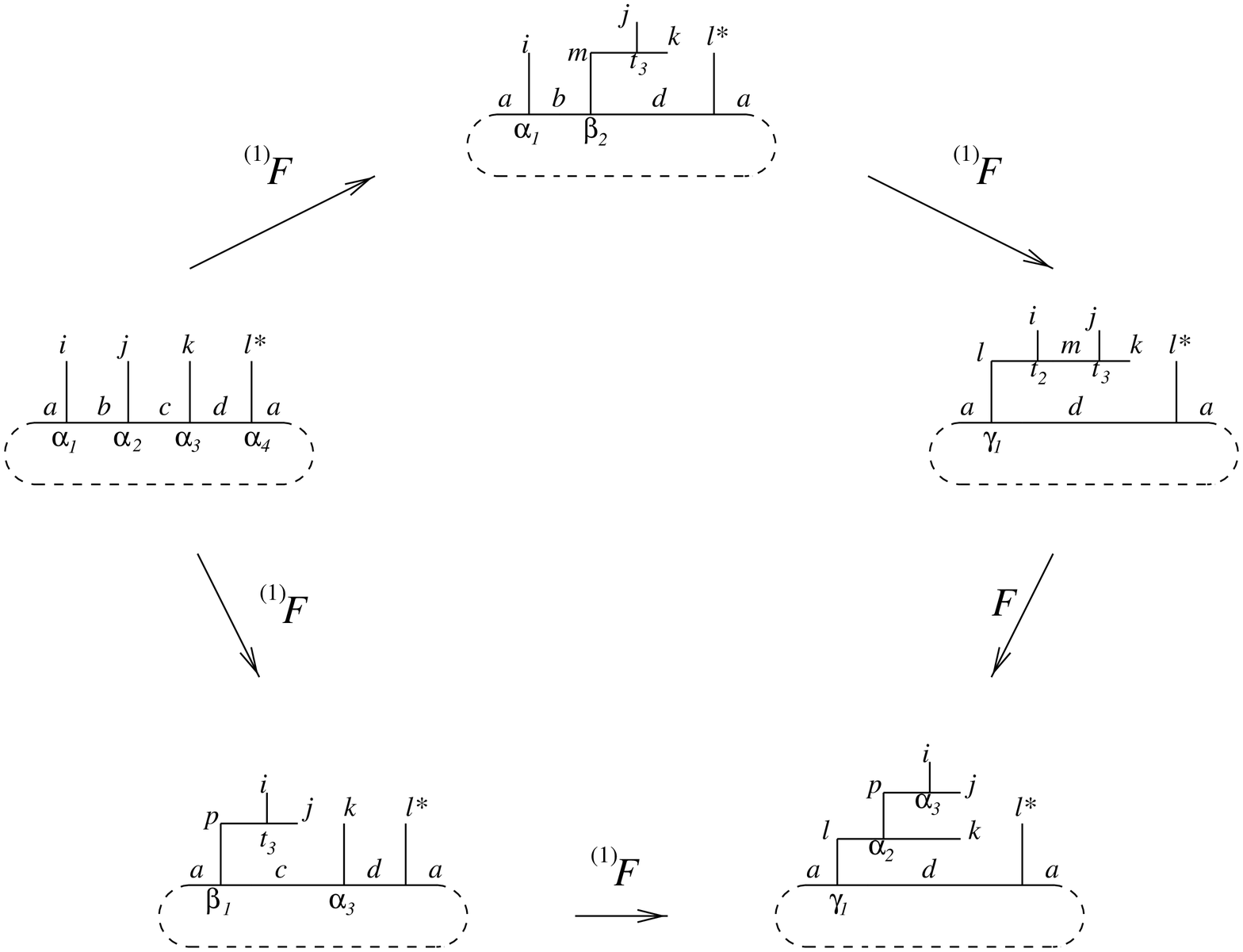}{11.0truecm}\figlabel\pentagon
It is thus a non trivial task to determine $\Fo$ for a given graph:
it may fail, in certain cases, because of some obstruction 
not revealed by the study of the set $\{n_i\}$ alone. 
This is what has been announced by Ocneanu \AO\ for one of
the graphs of $\slh(3)$ listed in \BPPZ.

\newsec{Generalised twisted boundary conditions }
\subsec{Twist operators}
\noindent
In the discussion of sect 1.2, we computed the partition function 
on a torus by identifying the states at the two ends of a cylinder
through the trace operation. Let us now imagine that we allow the 
action of a (non-local) operator $X$, attached to a cycle 
of the cylinder, before identifying these two 
ends, and thus compute $Z_X=\tr_{\CH_P} X e^{2\pi i[\tilde\tau(L_0-c/24)
-\tilde\tau^*(\bar L_0-c/24)]}$. 
The effect of this operator is to ``twist'' the b.c. In statistical
mechanics, this is a familiar operation, often referred to as a 
``seam''. A typical example is offered by the Ising model, where
antiperiodic b.c. may be imposed on the spins in the ``time'' direction
by inserting the operator $X$ which flips the spins along a 
generatrix of the cylinder. 
The $X$ are not arbitrary: we demand that they commute with the 
action of the two copies of the Virasoro algebra
\eqn\Xcomm{[L_n,X]=[\bar L_n,X]= 0 \ .}
As the $L_n$, $\bar L_n$ generate infinitesimal diffeomorphisms, 
this condition may be regarded as expressing that the operators 
$X$ are invariant under an arbitrary deformation of the closed 
line to which they are attached. Returning to our example of the 
lattice Ising model, the antiperiodicity condition on the spins
may be imposed along any non contractible, non self-intersecting
 curve around the cylinder, the deformation of which is made possible 
by the invariance of the model under local changes of the spin 
variable $\sigma_i \to -\sigma_i$. In general, when the CFT has a 
chiral algebra larger than Vir, we impose conditions of commutation 
of $X$ with all the generators of that algebra. 

\medskip
Now, the solutions of \Xcomm\ are readily found, through a line of 
argument parallel to that of sect. 2.2 and a new application of
Schur's lemma. $X$ acts in $\CH_P$ (see eq \hilbert) and is in general
a linear  combination of solutions of \Xcomm\ mapping some 
$\CV_j\otimes \overline{\CV_{\bj}}$  into $\CV_{j'}\otimes 
\overline{\CV_{\bj'}}$. 
Irreducibility of the $\CV$s tells us that $X$ is non trivial only 
if $j=j'$ and $\bj=\bj'$. If the multiplicity $Z_{j\bj}=1$,  
$X$  is then proportional to the projector in $\CV_j\otimes 
\overline{\CV_{\bj}}$.
If $Z_{j\bj}>1$,  however, $X$ is a linear superposition of operators
intertwining the different copies of  $\CV_j\otimes \overline{\CV_{\bj}}$ 
and acting as a projector in each
\eqn\schur{
P^{(j,\bj;\za,\za')}\ : \ (\CV_j\otimes  \overline{\CV}_{\bj})^{(\za')} 
\to (\CV_j\otimes \overline{\CV}_{\bj})^{(\za)} \quad \za,\za'=
1,\cdots, Z_{j\bj}\ .}
In other words, if $|j,\bn\ket\otimes |\bj,\bar\bn\ket$ is an 
orthonormal basis of  $\CV_j\otimes \overline{\CV_{\bj}}$, with $\bn,\bar\bn$ 
multi-indices labelling the descendent states,
\eqn\projjbj{P^{(j,\bj;\za,\za')}=\sum_{\bn,\bbn} 
\big(|j,\bn\rangle \otimes |\bj,\bbn\rangle\big)^{(\za)}
\big(\langle j,\bn| \otimes 
\langle\bj,\bbn|\big)^{(\za')}\quad \za,\za'=1,\cdots
Z_{j\bj} \ .}
There are thus $\sum_{j,\bj} \big(Z_{j\bj}\big)^2$ linearly independent 
operators $P^{(j,\bj;\, \za,\za')}$, 
which play in this problem the r\^ole of the Ishibashi states in 
the discussion of sect 2. They may be composed according to
\eqn\compo{P^{(j_1,\bj_1;\za_1,\za_1')}P^{(j_2,\bj_2;\za_2,\za'_2)}=
\zd_{j_1j_2}\zd_{\bj_1\bj_2}\zd_{\za_1'\za_2}\, P^{(j_1,\bj_1;\za_1,\za_2')}
\ .}
The most general solution $X$ of \Xcomm\ may then be written as
\eqn\Xlin{X_x= \sum_{j\bj,\za,\za'} 
{\Psi_x^{(j,\bj;\za,\za')}\over\sqrt{S_{1j}
S_{1\bj}}}\,  P^{(j,\bj;\za,\za')}\ ,}
where the $\Psi$ are complex numbers. 
The denominator $\sqrt{S_{1j} S_{1\bj}}$ is introduced for later
convenience. We shall denote by ${\tCV}$ the set of labels $x$ and  
use the label $x=1$ for the identity operator
\eqn\idenX{X_1:={\rm Id}=\sum_{j\bj,\za} P^{(j,\bj;\za,\za)}\ ,}
for which 
\eqn\Psiid{\Psi_1^{(j,\bj;\za,\za')}= \sqrt{S_{1j} S_{1\bj}}\,\delta_{\alpha
\alpha'}=: \Psi_1^{(j,\bj)}\, \delta_{\alpha \alpha'}  \ .}
Using \compo\ and  the hermitian conjugation properties of the projectors
\eqn\hermit{
( P^{(j,\bj;\za,\za')})^\dagger= P^{(j,\bj;\za',\za)}\, }
we may compose two such $X$  as
\eqn\Xcomp{
X_x^\dagger\, X_y= \sum_{j,\bj,\za,\za',\za''}{\Psi_x^{(j,\bj;\za,\za')\, *}\ 
\Psi_y^{(j,\bj;\za'',\za')} \over{S_{1j} S_{1\bj}}}\,
 P^{(j,\bj;\za,\za'')}\ .} 
Insertion of one or two such $X$ will be sufficient to expose their 
most interesting features. 

\subsec{A new consistency condition}
\noindent
We may now repeat the steps followed in sect 2. We consider a 
finite cylinder of length $2L$ and of circumference $T$. If 
$\tilde q=\exp 2i\pi \tilde\tau$, $\tilde\tau=2iL/T$, we have
\eqn\char
{\tr_{\CH_P}( P^{(j,\bj;\za,\za')}
\tq^{L_0-c/24}\, \btq^{\bL_0-c/24})
 = \chi_j(\tq)\, \chi_{\bj}(\btq)\  \delta_{\za\za'}\ ,} 
and the partition function on that cylinder in the presence of 
two twists reads
\eqn\Zxy
{Z_{x|y}:=Z_{X_x^\dagger\,X_y}= \sum_{j,\bj\in \CI\atop \za,\za'=1,\cdots,Z_{j\bj}} 
{\Psi_x^{(j,\bj;\za,\za')\, *}
\Psi_y^{(j,\bj;\za,\za')}\over S_{1j}S_{1\bj}} \
 \chi_j(\tq)\, \chi_{\bj}(\btq)\,.
}
In particular, for $x=y=1$, we find
\eqn\modinv{Z_{1|1}= \sum_{j,\bj,\za} 
\chi_j(\tq)\, \chi_{\bj}(\btq)=
\sum_{j,\bj\in \CI}\, Z_{j\bj}\, \chi_j(\tq)\, \chi_{\bj}(\btq)\ ,}
which is the modular invariant partition function
describing the system with no twist.

As the commutation relation \Xcomm\ guarantees the existence 
of a well defined stress-energy tensor on the cylinder consistent 
with the identification of its two ends, 
we may also carry out the computation of $Z_{x|y}$ in another way, 
by making use of the 
Hamiltonian corresponding to the $\Re e\, w$-translation operator.
Then the partition function $Z_{x|y}$ is obtained  
as the trace of the corresponding evolution operator in a  Hilbert space
\eqn\hilbertxy
{ \CH_{x|y}=\oplus_{i,\bi\in\CI}\, \tV_{i\bi^*;\, x}{}^y\, \CV_i\otimes
\overline{\CV}_{\bi}\ ,}
where the non negative integer multiplicities $ \tV_{i\bi;\, x}{}^y$ depend
on the twists $x$ and $y$. 
In the trivial case $x=y=1$, they must reduce to 
\eqn\trivial{ \tV_{i\bi^*;\, 1}{}^1=Z_{i\bi}\ .}
We can thus complete the calculation as in the
absence of the $X$ operator(s) and get, with $q=\exp 2\pi i \tau$, 
$\tau=-1/\tilde\tau=iT/2L$
\eqnn\Zpxy
$$\eqalignno{
Z_{x|y}&=\tr_{\CH_{x|y}}\, q^{L_0-c/24}\,
 q^{\bL_0-c/24}\,  \cr
&=  
\sum_{i,\bi\in\CI} \tV_{i\bi;\, x}{}^y \ \chi_i(q)\, \chi_{\bi}(q)\ .
&\Zpxy }
$$

Identifying the two expressions \Zxy\ and \Zpxy\ after a modular
transformation of the characters, we get
\eqn\Cardy
{  \tV_{i\bi;\, x}{}^y= \sum_{j,\bj,\za,\za'}\, 
{S_{ij}S_{\bi\bj} \over S_{1j}S_{1\bj}}\
\Psi_x^{(j,\bj; \za,\za')}\ \Psi_y^{(j,\bj;\za,\za')\, *}
\,,\qquad i,\bi\in\CI \,.
}
In fact, as in sect 2, there is an ambiguity inherent to the use of
specialised characters. See \PZtw\ for a discussion of that point. 
The similarity of condition \Cardy\ with Cardy's equation in the case
of open boundaries is quite striking. We shall in fact 
exploit  equation \Cardy\ in a way parallel to what we did in sect 2.

To proceed, we make the additional assumption that the 
$\Psi_x^{(j,\bj;\, \za,\za')}$ form a unitary (i.e. orthonormal 
and complete) change of basis  
from the $P^{(j,\bj,\za,\za')}$ to the $X_x$ operators. 
This implies that the set $\tCV$ of labels $x$ has a cardinality equal
to $\sum (Z_{j\bj})^2$. 
The integer numbers $\tV_{i\bi;x}{}^y$ will be considered either as 
the entries
of $|\CI|\times|\CI|$ matrices $\tV_x{}^y$, $x,y\in\tCV$,  or as those
of $|\tCV|\times |\tCV|$ matrices $\tV_{i\bi}$, $i,\bi\in\CI$. 

Following the same argument as in sect 2.4, 
equation \Cardy\ may be regarded as
the spectral decomposition of the matrices $\tV_{i\bi}$ into their
orthogonal eigenvectors $\Psi$ and eigenvalues $S_{ij}S_{\bi\bj}/
S_{1j}S_{1\bj}$. As the latter form a representation 
of the tensor product of two copies of Verlinde fusion algebra, the same
holds true for the $\tV$ matrices:
\eqn\doublefus
{ \tilde{V}_{i_1j_1} \tilde{V}_{i_2j_2}=
\sum_{i_3,j_3} \CN_{i_1i_2}{}^{i_3} \CN_{j_1j_2}{}^{j_3}\
\tilde{V}_{i_3j_3}\ .}
Combining \trivial\ with \doublefus, we have in particular
\eqn\Mx{
\sum_{i_3j_3} \CN_{i_1i_2}{}^{i_3} \CN_{j_1j_2}{}^{j_3}\, Z_{i_3j_3}=\sum_x
\tV_{i_1j_1^*;\, 1}{}^x \tV_{i_2j_2^*;\,x}{}^1 \ , }
which is the way the matrices $\tV_{ij;\, 1}{}^x=\tV_{i^*j^*;\, x}{}^1$
 appeared originally in the work of Ocneanu.
As will be explained below, all $\tV_x{}^y$ may be reconstructed
from  the simpler Ocneanu matrices $\tV_1{}^x$.
According to an argument already used for the $n$ matrices, 
by \doublefus, it is sufficient to specify $\tV_{f1}$ and $\tV_{1f}$ 
 for a generating subset of representations $f\in\CI$
to determine the whole set of matrices $\tV_{ij}$;
for example $f=2$ in the 
case of $sl(2)$ theories. The matrices  $\tV_{f1}$ and $\tV_{1f}$ 
 may be regarded as the adjacency matrices of graphs with the 
common set of vertices $\tCV$. 	It is convenient to draw these graphs
on the same chart, with edges of different colours, and we refer 
to the resulting multiple graph ${\tilde G}$ as Ocneanu graph. 

\fig{The Ocneanu graph of type $E_6$: the edges corresponding
to $\tV_{21}$, resp. $\tV_{12}$, have been drawn in full red, 
resp. blue broken}{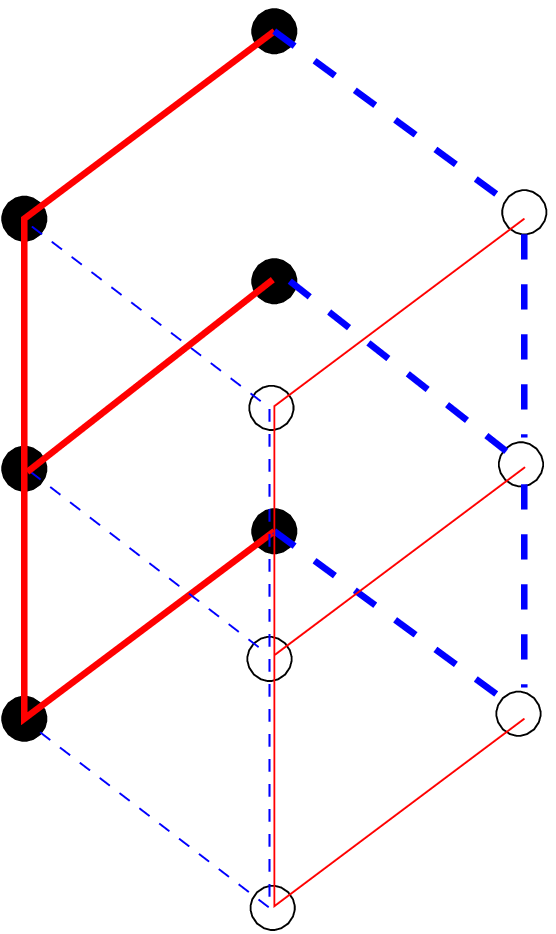}{2.5truecm}\figlabel\ocnesix
In diagonal cases, 
the matrices $\tV_{ij}$ are labelled by the set
$\tCV=\CI$ of representation labels and must satisfy \doublefus\ 
and \trivial. They are simply
\eqn\tVdiag{\tV_{ij}=\CN_i\CN_j\ ,}
i.e. $\tV_{ij;\,x}{}^y=\sum_k \N_{ix}{}^k \N_{kj}{}^y$, 
and the Ocneanu graph is simply made of two identical copies of 
the fusion graph of $\CN_f$ with their vertices identified. 
The example of the Ocneanu graph for 
the $\slh(2)$ theory of type $E_6$ is presented on Fig. \ocnesix. 
More graphs and more details on these graphs and their 
construction may be found in \refs{\Ocn,\AO,\Xu,\BE,\BEK,\PZnew}.

\subsec{The fusion algebra of defect lines}
\noindent
We may finally combine the two situations encountered in the 
discussion of these lectures and consider the insertion of twist operators
in the presence of open boundaries with boundary states $|a\ket$
and $\bra b|$. It is a  simple exercise left to
the reader to check that 
\eqn\Zaxb{Z_{ax}{}^b:=\bra b| X_x^\dagger \tq^{L_0-{c\over 24}}  
\tbq^{\bar L_0-{c\over 24}}  |a\ket=\sum_i 
(n_i\tilde{n}_x)_{a^*}^{\ b^*} \chi_i(q)\ , }
with $\tn_x=\{\tn_{ax}{}^b\}$ a new set of multiplicities, 
i.e. non negative integers, 
\eqn\tnm{
\tilde{n}_{ax}{}^b=\sum_{j,\;\za, \zb }\, \psi_a^{(j\,,\za)}\,
{\Psi_x^{(j, j;\, \za,\zb)}\over \Psi_1^{(j,j)}}\,
\psi_b^{(j\,,\zb)\, *}\, .}
These new matrices $\tn_x$ play a r\^ole parallel to that of the $n_i$, 
and indeed, they form themselves a (non-negative integer valued 
matrix) representation of an algebra with (non negative integer)
structure constants
\eqn\nimreptn{\tn_x\tn_y=\sum_z \tN_{xy}{}^z \tn_z\ ,}
\eqn\Ibl{
\tN_{x y}{}^z=\sum_{j,\bj;\za}\,\sum_{\zb,\zg }\, \Psi_x^{(j,\bj;\,\za,\zb)}\,
{\Psi_y^{(j,\bj;\, \zb,\zg)}\over \Psi_1^{(j,\bj)}}\,
\Psi_z^{(j,\bj;\, \za,\zg)\, *}\,.
}
One also easily checks that 
\eqn\fustw{ Z_{y|z}=\sum_x\, \tN_{yx}{}^z\ Z_{1|x}\ ,}
or equivalently, $\tV_y{}^z= \sum_x \tN_{yx}{}^z \tV_1{}^x$, as announced 
above. The matrices $\tN_x:=\{\tN_{yx}{}^z\}$ form an associative
algebra 
\eqn\deffa{\tN_x\tN_y=\sum_z \tN_{xy}{}^z\tN_z\ ,} 
which may be called the ``fusion algebra of defect lines''.

There is, however, a major and most intriguing new feature which
distinguishes this new algebra $\tN$ from the standard fusion algebra
$N_i$. In general, if some of the multipicities $Z_{j\bj}>1$, 
this new algebra is non commutative! \Ocn. Physically, this must be 
interpreted as the lack of commutativity of the twist operators 
(or of the defect lines to which they are attached). 

We conclude this section with the remark that the Ocneanu graphs
$\tilde{G}$ contain an information on the  OPA of fields with
arbitrary spins, generalising  the relation of the scalar fields
OPE coefficients to the Pasquier algebra discussed above, see
\PZnew\ for details.

\fig{The simplices}{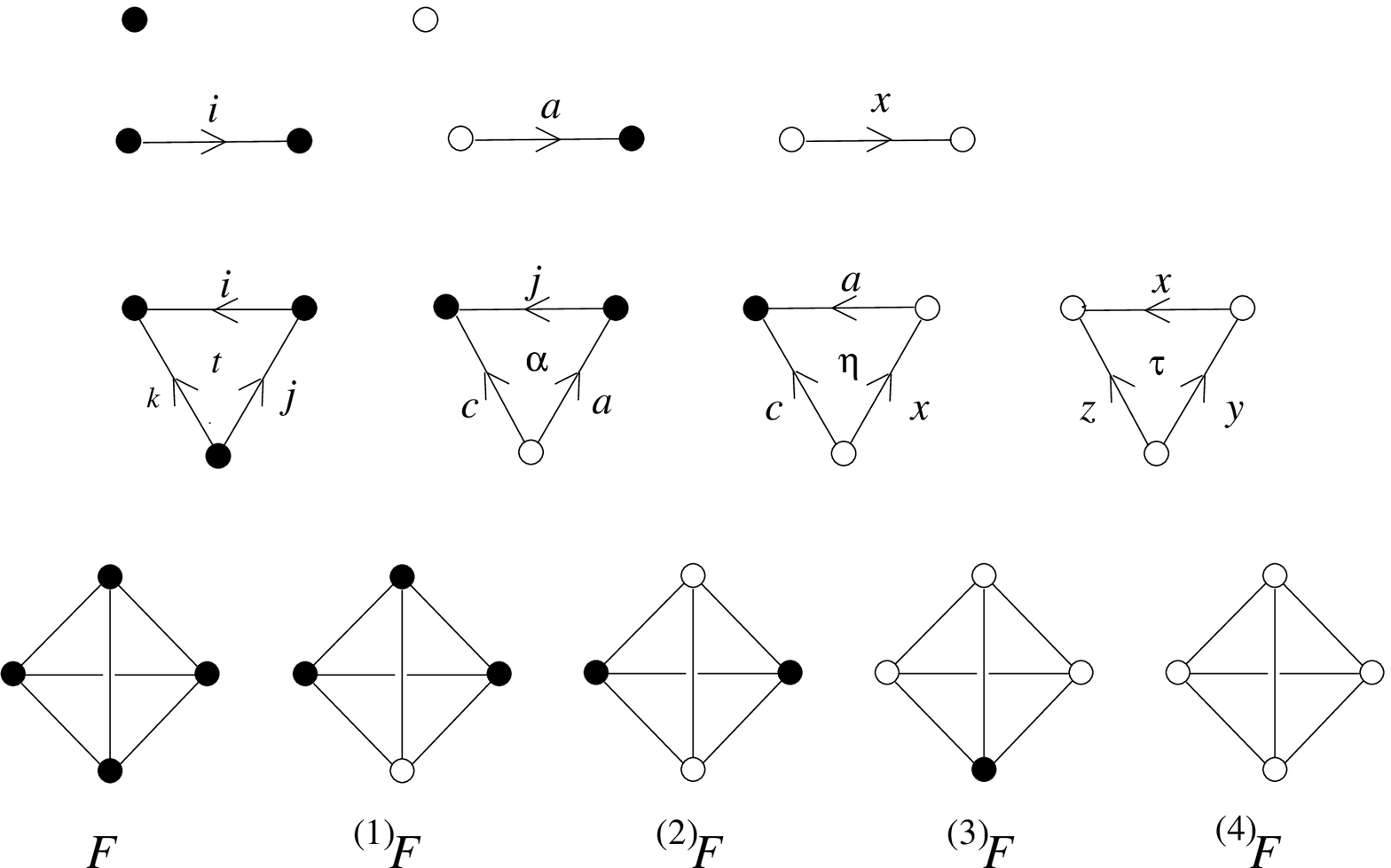}{9cm}\figlabel\simplex

\newsec{The underlying quantum algebra}

\noindent
The various sets of multiplicities that we have encountered
$\{\CN_i\}$, $\{n_i\}$, $\{\tN_x\}$ and $\{\tn_x\}$, and the 
inter-relations that they satisfy, 
are in fact the indicators of a deep and beautiful structure.
In the combinatorial approach of \refs{\Ocn,\BSz}, one 
constructs a simplicial complex out of the following elements (Fig
\simplex): 2 kinds of 0-simplices, indicated by black or white
dots, 3 kinds of oriented edges, 4 kinds of 2-faces and 5 kinds 
of tetrahedral 3-simplices. The triangular faces come with a 
multiplicity index, $t=1,\cdots,\CN_{ij}{}^k$, $\za=1,\cdots,n_{ia}{}^c$, 
$\eta=1,\cdots, \tn_{ax}{}^c$, $\tau=1,\cdots,\tN_{xy}{}^z$,
and these multiplicities obey the set of relations 
\eqnn\rela
$$\eqalignno{
\sum_{\ell\in \CI}\CN_{im}{}^\ell \CN_{j\ell}{}^n 
&= \sum_{k\in \CI}\,\CN_{ij}{}^k\, \CN_{km}{}^n & \regfa\cr
\sum_{c\in \CV} n_{ia}{}^c n_{jc}{}^b 
&= \sum_{k\in \CI}\,\CN_{ij}{}^k\, n_{ka}{}^b & \nfus \cr
\sum_{x\in\tCV} \tn_{ax}{}^{a'}\tn_{x^* b'}{}^b&=
\sum_{i\in\CI} n_{ia}{}^b\,n_{i^* b'}{}^{a'}  & \rela \cr
\sum_{c\in \CV}\tn_{ax}{}^c\tn_{cy}{}^b
&=\sum_{z\in \tCV} \tN_{xy}{}^z \tn_{az}{}^b & \nimreptn \cr
\sum_{v\in \tCV}\tN_{wx}{}^v\tN_{vy}{}^u
&=\sum_{z\in\tCV} \tN_{xy}{}^z \tN_{wz}{}^u\ . & \deffa \cr}$$
\fig{Pentagonal identity resulting from two ways of cutting a solid
into  tetrahedra}{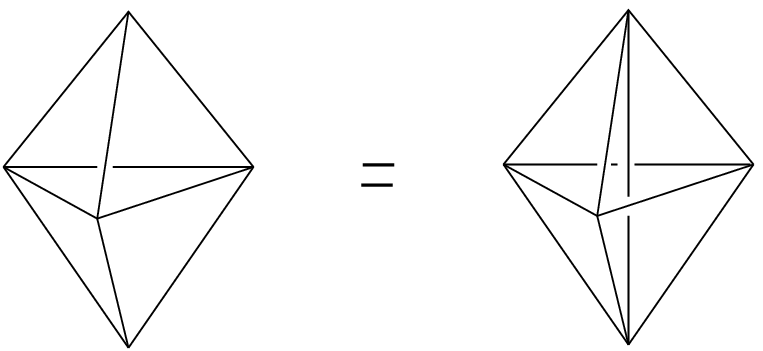}{5cm}\figlabel\doubletr
With each tetrahedron, we want to attach a complex number, (a 3-chain), 
$F_{m n}\!\!\left[{i\atop k}{j\atop l}\right]_{t  t'}^{u u'}\,$, 
$\Fo_{b k}\!\!\left[{i\atop c}{j\atop a}\right]_{\za\za'}^{\zb t}\,$, 
${{}^{(2)}\!F}_{ix}\!\!\!\left[{a\atop c}{b\atop d}\right]_{\za\zb}
^{\eta\zeta}\,$, 
${{}^{(3)}\!F}_{bz}\!\!\left[{x\atop c}{y\atop a}\right]_{\eta\eta'}
^{\zeta \tau}\,$, 
${{}^{(4)}\!F}_{uv}\!\left[{x\atop w}{y\atop z}\right]_{\tau\tau'}
^{\upsilon\upsilon'}\,$,
which must satisfy all the pentagon identities resulting from the 
decorations with black and white colors of the vertices of the 
solid of Fig \doubletr, and from its cutting either into three 
or into two tetrahedra. These identities include those already 
mentioned in  sect 1.1 and 3. As in these previous encounters, 
they realise 
 unitary (with the exception of the invertible ${{}^{(2)}\!F}$)
changes of basis, and the five relations above are necessary
conditions for each pair of bases,
(depicted by double triangles with a fixed $0,1,2,3,4$ number of
white vertices respectively) to have the same dimension. 

These data, if they exist, enable one to construct an abstract 
algebra, the Ocneanu ``double triangle algebra'' \AO,
which appears as the quantum symmetry of the CFT \PZnew. Being defined
in terms of the $n$'s etc, this algebra is intrinsic to the CFT, 
its spectrum of bulk and boundary states etc. 
We refer the reader to \refs{\AO,\BSz,\BEK,\PZnew}
 for a thorough discussion of its definition, its various 
interpretations, and for a fairly extensive list of references. 

{\petit The algebra $\CA$ is constructed as follows. One first 
considers a basis $|e_{ba}^{j,\za}\rangle$ labelled and represented 
pictorially in the same way as the boundary fields of sect 3,
or by the triangles with one white vertex  in Fig. 6.
 One defines tensor products of two such vectors provided the 
intermediate labels coincide,  and the matrix elements of $\Fo$ 
act as ``$3j$-symbols'', while the $F$ are the recoupling,
or $6j$-, symbols. The algebra itself is generated by the 
matrix units $|e^{j,\zb}_{ca}\ket\bra e^{j,\zb'}_{c',a'}|$. 
One shows that this algebra has a  product and a 
coassociative coproduct 
(inherited from the previous tensor product),
a counit and an antipode, but that the axioms 
of Hopf algebras are not all satisfied: $\CA$ is a ``weak Hopf 
algebra'', also called a ``quantum groupoid'' \BSz. 
Its dual has a basis labelled by indices $x$ as above, relation
\rela\ guarantees 
the existence of an invertible change ${{}^{(2)}\!F}$ from the original 
basis to the dual one. In the dual, the matrices ${{}^{(3)}\!F}$
and ${{}^{(4)}\!F}$ play the r\^oles of $3j$- and $6j$-symbols.
The dimension of $\CA$ is given by the number of all double triangles
with two black and two white vertices, i.e., by the sum over $a,a',b,b'$
of each side of \rela.}

\newsec{Integrable lattice models}

\nind 
It should be recalled that in parallel to the conformal
field theoretic discussion sketched in these notes, one may  
study  a class of lattice integrable models, the so-called
face, or height, or RSOS, models, which are also described in terms
of the same graphs. There the  degrees of freedom are attached to
sites of a square lattice, and are assigned to take their value 
in the set of  vertices of the chosen graph. Typically, in the 
simplest models, neighbouring sites on the lattice are assigned   
neighbouring vertices on the graph. Boltzmann weights are
given for each configuration of four vertices around a square face.
They depend on an additional parameter, the spectral parameter $u$, 
and must satisfy an integrability condition, 
the celebrated  Yang-Baxter equation. This is  realised algebraically 
through a representation of the Temperley-Lieb algebra, or of some
other quotient of the Hecke algebra, constructed  
on the graph as follows. 
\def\bn{{\bf n}}
For a triplet of sites $\bn-1,\bn,\bn+1$ along 
a diagonal zig-zag line on the lattice, and ``heights'' $a$, $b$ or $d$ 
and $c$ assigned to them, the face Boltzmann weight reads 
\eqn\facew{
 X_{{\bf n}}(u)=\hbox{\raise -9.mm\hbox{\epsfxsize=1.5cm\epsfbox{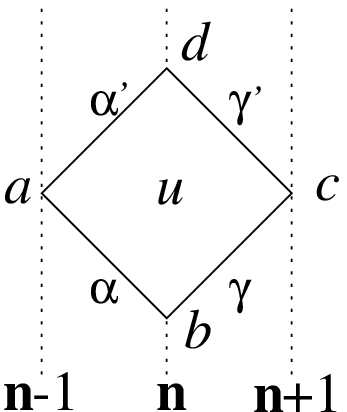}}}
=\sin\({\pi\over h}-u\)\delta_{bd}+\sin u\,\(U_{\bn}\)_{bd}}
The $U$ are the generators of the Hecke algebra, i.e.  satisfy 
\eqnn\hecke
$$\eqalignno{U_\bn^2 &= 2\cos {\pi\over h}\ U_\bn\cr
U_{\bn}U_{\bn+1}U_{\bn}-U_{\bn}&= U_{\bn+1}U_{\bn}
U_{\bn+1}-U_{\bn+1} & \hecke\cr
U_{\bf n} U_{\bf m}&= U_{\bf m} U_{\bf n} \qquad {\rm if}\ |{\bf n}-{\bf m}|
\ge 2 \ . 
}$$
As a consequence,  the face weights satisfy the Yang-Baxter identity
\eqn\yb
{ X_\bn(u) X_{\bn +1}(u+v) X_\bn(v) = X_{\bn +1}(v)X_\bn(u+v) X_{\bn+1}(u)\ .}
The Pasquier models \VPun\ give the simplest and most explicit example: 
the relevant graphs are once again
the ADE diagrams, and the $U$ matrix element  reads
\eqn\pas{ (U_\bn)_{bd}  
= \delta_{ac}\,{\sqrt{\psi^1_b\,\psi^1_d}\over\psi^1_a }\ .}
\def\youngone{{ \hbox{\epsfxsize=1.5mm\epsfbox{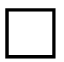}} }}
They are related to the $sl(2)$ algebra. 
Although  generalisations to higher rank are known to exist, 
there is a  lack of such  explicit and general formulas. 
In general, there are  additional edge degrees of freedom
$\alpha,\gamma,\alpha',\gamma'$, $\alpha=1,\cdots,n_{\youngone a}{}^b$, etc.
The weights for the (generalised) $A$-type graphs are known 
from the work of Jimbo et al, and  Wenzl \JW. 
In \refs{\DFZun,\Soch} explicit expressions have been given 
for models associated with some of the graphs of $\slh(3)$, 
and a general result has been obtained in \Xu\ for graphs of $\slh(N)$ 
corresponding to conformal embeddings. The  recent 
observation by Ocneanu \AO\ that one may write the above generator
of the Hecke algebra as
\def\youngtwo{ \hbox{\raise
-0.5mm\hbox{\epsfxsize=1.5mm\epsfbox{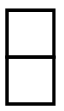}} }} 
\eqn\faceocn
{ {U_{bd}}_{\gamma\alpha}^{\gamma'\alpha'}=\sum_{\zb}\,
\Fo_{b\, \youngtwo}\left[{\youngone\atop c} {\youngone\atop
a}\right]_{\zg\,
\za}^{\zb\ 1}\, \Fo_{d\, \youngtwo}^*\left[{\youngone\atop c}
{\youngone\atop a}
\right]_{\zg'\, \za'}^{\zb\ 1 }\,
}
in terms of the $\Fo$ matrix previously introduced is thus a
significant progress, both practically and conceptually, 
since it connects problems of apparently different nature.  

\medskip
Finally, in these lattice models, it is legitimate to
wonder if boundaries may be introduced without spoiling integrability.
This requires  a careful
determination of the boundary Boltzmann weights, satisfying the
so-called Boundary Yang-Baxter Equation \BP.  This has now been 
completed for the unitary minimal models: a large class of
boundary weights has been found, which at criticality, match
perfectly what we have learnt from BCFT \BPde.  
It remains to connect these boundary weights with quantities defined 
previously in the context of BCFT to have a fully consistent and 
unified picture of all questions of boundary conditions in integrable
lattice models and conformal theories.


\bigskip\noindent{\bf Acknowledgements}\par\noindent
J.-B. Z. is happy 
to thank Professors Z. Horvath and L. Palla,   
 the E\"otv\"os University and Bolyai College
for their invitation and hospitality in the wonderful city of Budapest
and all the participants of the school and conference
for providing a stimulating atmosphere. 
A good part of the work presented here results from several  
enjoyable collaborations  with R. Behrend and P. Pearce.  


\vskip1cm

\footatend\vfill\supereject\immediate\closeout\rfile\writestoppt
\baselineskip=14pt
\noindent{{\bf  References}}\bigskip{\ninerm\frenchspacing%
\parindent=20pt\escapechar=` \input refs.tmp\vfill\eject}\nonfrenchspacing

\bye